\documentclass{article}%
\usepackage{amssymb}
\usepackage{amsfonts}
\usepackage{amsmath}%
\usepackage{graphicx}
\providecommand{\U}[1]{\protect\rule{.1in}{.1in}}

\begin{document}

\title{Quantum Probability Explanations for Probability Judgment `Errors'}
\author{Jerome R. Busemeyer\\Indiana University, USA
\and Riccardo Franco\\Italy
\and Emmanuel M. Pothos\\Swansea University, UK}
\maketitle

\begin{abstract}
A quantum probability model is introduced and used to explain human
probability judgment errors including the conjunction, disjunction, inverse,
and conditional fallacies, as well as unpacking effects and partitioning
effects. Quantum probability theory is a general and coherent theory based on
a set of (von Neumann) axioms which relax some of the constraints underlying
classic (Kolmogorov) probability theory. The quantum model is compared and
contrasted with other competing explanations for these judgment errors
including the representativeness heuristic, the averaging model, and a memory
retrieval model for probability judgments. The quantum model also provides
ways to extend Bayesian, fuzzy set, and fuzzy trace theories. We conclude that
quantum information processing principles provide a viable and promising new
way to understand human judgment and reasoning.

\end{abstract}

Over 30 years ago, Kahneman and Tversky \cite{T&K_Science_HB_1974} began their
influential program of research to discover the heuristics and biases that
form the basis of human probability judgments. Since that time, a great deal
of new and challenging empirical phenomena have been discovered including
conjunction, disjunction, conditional, inverse, and base rate fallacies
\cite{Gilovich_HB_2002}. Although heuristic concepts (such as
representativeness and availability) initially served as a guide to
researchers in this area, there is a growing need to move beyond these
intuitions, and develop more coherent, comprehensive, and deductive
theoretical explanations \cite{ShahOppenheimer_2008}. The purpose of this
article is to propose a new way of understanding human probability judgment
using quantum probability principles \cite{GudderQP}. \ Quantum principles
have been used recently in a number of psychological applications including
perception \cite{Atmanspacher_BioCyb_2004}, conceptual structure
\cite{Aerts&Gabora_2005}, information retrieval \cite{VanRijsbergen_2004}, and
human judgments \cite{KhrenBook}.\footnote{There is another independent line
of research that uses quantum physical models of the brain to understand
consciousness \cite{Hameroff_1998} and human memory \cite{Pribram_1993}. We
are not following this line, and instead we are using models at a more
abstract level analogous to Bayesian models of cognition.}

The concept of a probability judgment error requires a standard or norm, and
in the past, this norm was based on the Kolmogorov axioms for classic
probability theory \cite{Kolmogorov}. Classic theory is based on the
assignment of probabilities to events defined as sets, and the Boolean logic
entailed by using sets seems to be the source of the problems that occur with
applications to human judgments. Quantum probability provides a more general
geometric approach to probability theory that remains coherent but relaxes
some of the constraints of Boolean logic \cite{Pitowski_1989}. Thus quantum
probability provides an opportunity to explain what appears to be judgmental
`errors' with respect to the classic definition, but at the same time, it
provides a quantum logical `rationale' for human probability judgments.

The remainder of this article is organized as follows. First we provide some
background information and review basic findings. Second, we provide a simple
and elementary introduction to quantum probability theory and apply these
ideas to the basic findings. Finally, we summarize previous theoretical
explanations, compare the advantages and disadvantages of the quantum model
with the previous models, and indicate directions for future research.

\section{Background and Brief Review}

This article is mainly concerned with the explanation of conjunction and
disjunction fallacies, and so the following review and later theoretical
analyses focus on these two basic issues. However, it is important to briefly
examine how well this explanation extends to some closely related phenomena,
including conditional and inverse fallacies and `unpacking' effects.
Therefore, although we focus on conjunction and disjunction fallacies, we also
briefly examine some closely related fallacies. This review addresses the many
qualitative (ordinal level) findings that have been discovered over the past
30 years.

In many probability judgment studies, a story is provided which is followed by
questions about the likelihood of events related to the story (e.g., a story
about a liberal philosophy student from Berkeley named Linda is presented, and
questions are asked about her future activities). Sometimes very little story
is needed (e.g. a time and a place) and there is simply a causal connection
between story events (e.g., an increase in cigarette tax is passed, and then a
decrease in teenage smoking occurs). Some of the key experimental factors that
are manipulated in these studies include the following. Questions about events
can be \textit{related} by referring to the same person (e.g., `Linda is a
bank teller', `Linda is active in feminist movement') or \textit{unrelated} by
referring to different people (`Linda is active in feminist movement', `Bill
is shy'). \ Questions about events can have \textit{high} likelihood (e.g.,
`Linda is active in feminist movement') or a \textit{low} likelihood (`Linda
is a bank teller'). \ Questions can be about events with positive (e.g., `Bill
enjoys jogging and Bill plays soccer ') or negative or zero dependencies
(e.g., Bill is an accountant and Bill likes jogging').

Questions about generic events are labeled by letters such as $A$ and $B$. We
use the letters $H$ and $L$ to denote questions about events that have a high
or low likelihood, respectively. Sometimes, subscripts on the letters will be
used to distinguish questions about events that are related or unrelated. For
example, $A_{1}$ and $B_{1}$ refer to events that are \textit{related} (e.g.,
Tei has blue eyes, Tei has blond hair); $A_{1}$ and $B_{2}$ refer to events
that are \textit{unrelated} (e.g., Tei has blue eyes, Jerry has blond hair).
When no subscripts appear, it can be assumed that the events are related. The
probabilities of interest include questions about a single event (e.g., `is a
$A$ true?'), a negation of a question (`is not $A$ true?' symbolized as
$\symbol{126}A),$ a conjunctive question about events (`is $A$ and $B$ true?'
symbolized $A\wedge B$), a disjunctive question about events (`is $A$ or $B$
true?' symbolized $A\vee B$), and a question about an implication (`if $A$ is
true, then is $B$ true?', symbolized as $A\mapsto B$). The symbols $\wedge$
and $\vee$ represent the classic Boolean logic conjunction and disjunction
relations, which are commutative, $(A\wedge B)\leftrightarrow(B\wedge A)$ and
$(A\vee B)\leftrightarrow(B\vee A)$ and distributive $A\wedge(B\vee
\symbol{126}B)\leftrightarrow(A\wedge B)\vee(A\wedge\symbol{126}B).$ The
implication is not commutative $(A\mapsto B)\nleftrightarrow(B\mapsto A).$
These logical properties are intended by the experimenter asking the
questions, but they may not necessarily be treated this way by human judges
when answering questions about these events. Later, when various theoretical
explanations for the findings are presented, different symbols are used for
negation, conjunction, disjunction, and implication, because the formal
properties of these logical relations differ across theories.

Participants are asked to judge probabilities for questions about events, and
these judgments are denoted by the letter $J$. The judged probabilities
corresponding to the single, negation, conjunction, union, and implication
questions about events are denoted $J(A)\,$, $J(\symbol{126}A),$ $J(A\wedge
B)$, $J(A\vee B)$, and $J(A\mapsto B)$. These judgments may be obtained using
a choice response (e.g. which event is more likely), or rank ordering the
likelihood of a list of events, or rating each event (e.g. what are the
chances out of 100 that an event is true), and sometimes they are inferred
from bets (e.g. decide which event you want to bet money). To evaluate whether
or not a fallacy or judgment error occurs, one needs to compare the
distribution of judgments across participants for one event with another. This
is usually done using two methods: One is to compare the means (or medians) of
the two distributions and determine whether the difference is statistically
significant; the second is to compare the frequency of the correct versus
incorrect orders and determine whether the frequencies are statistically
different. These two methods usually but not always give the same answer when
they are both reported.

\subsection{Basic Findings}

\begin{figure}
[ptb]
\begin{center}
\includegraphics[
natheight=6.915900in,
natwidth=5.188900in,
height=4.7859in,
width=3.5985in
]%
{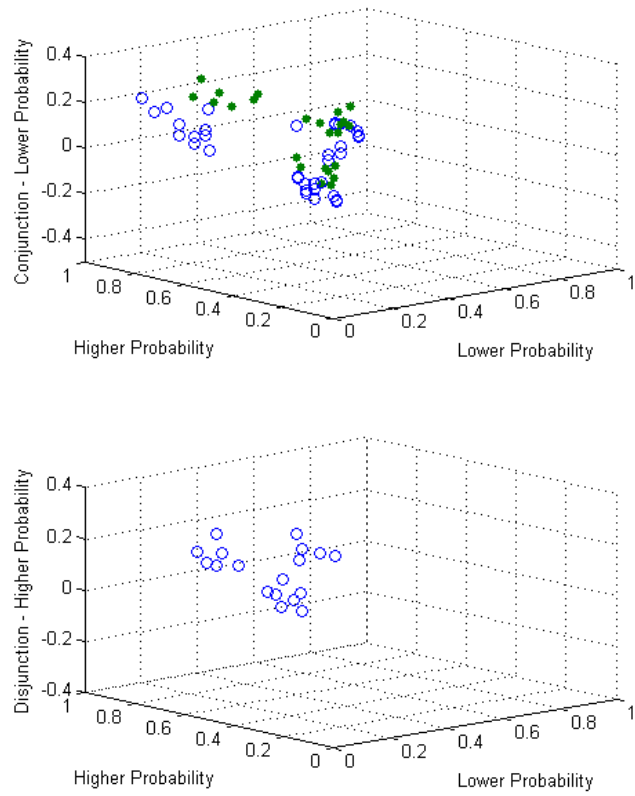}%
\caption{Top panel shows the mean conjunction effect from 60 experimental
conditions, and bottom panel shows mean disjunction effect from 18
experimental conditions.}%
\label{Gavansky_Fig}%
\end{center}
\end{figure}

As mentioned earlier, this article is primarily concerned with conjunction and
disjunction fallacies and some other closely related fallacies.\footnote{There
is a large literature on inference that we plan to address in future work, but
not at this time. In particular, we do not address the large literature on the
insensitivity to base-rates in Bayesian inference tasks
\cite{Koehler_baserate_1996}.} \ Figure \ref{Gavansky_Fig} provides a general
overview of (a) the magnitude of conjunction errors \cite{Gavanski_1991} in
the top panel and (b) the magnitude of disjunction errors \cite{Fisk_2002} in
the bottom panel.%

This figure plots the means of $J(H)$ and $J(L)$ along the X,Y axes, and the Z
(vertical) axis has the mean of $J(H\wedge L)-J(L)$ for the conjunction error
and the mean of $J(H)-J(H\vee L)$ for the disjunction error. The 36 open
circles ($N=40$ observations per circle) in the top panel are from Table 1 of
Gavansky \& Roskos-Ewoldsen (1991) in which people judged the conjunction
after the constituents; the 24 solid dots ($N=50$ observations per dot) in the
top panel are from Table 2 of Gavansky and Roskos-Ewoldsen (1991) in which
people judged the conjunction before the constituents, and the 18 circles
($N=88$ observations per circle) are from Experiment 2 of Fisk (2002) in which
people judged the disjunction and the constituents in a randomized order.
Points that lie above zero on the Z-axis indicate an error for the means. When
the conjunction was rated last, five large (greater than .10) conjunction
errors occurred and they all occurred for $J(L)<.3$ and $J(H)>.8$; when the
conjunction was rated first, 7 large (greater than .10) conjunction errors
occurred, and they all occurred for $J(L)<.40$ and $J(H)>.70$; 8 large
(greater than .10) disjunction errors occurred, and all but two occurred for
$J(L)<.30$ and $J(H)>.60$.\footnote{The two exceptions were $J(L)=.19$,
$J(H)=.42$, $J(H\vee L)=.29$ and $J(L)=.62$, $J(H)=.85$, $J(H\vee L)=.72$.} In
summary, large mean conjunctive and disjunctive errors tend to occur with a
high-low combination, they tend to disappear when $J(L)$ is approximately
equal to $J(H)$, and more errors occur when the conjunction is rated first as
compared to last. \ Next we consider how various other factors moderate these
effects, and we also review some other closely related probability judgment errors.

F1. Conjunctive fallacy: $J(H\wedge L)>J(L)$ \cite{T&K_Conj_1983}. This has
been found comparing means, medians, and frequencies. For example, when
presented the liberal Linda story, 85\% of 142 participants chose the event
`bank teller and feminist' as more likely than `bank teller' in a direct
choice between these two events. This high rate of conjunction errors persists
even when both conjunctions, $(H\wedge L)$ as well as $(H\wedge\symbol{126}%
L),$ are included in the list \cite{Wedell&Moro_2008}. Other examples include
a Norwegian student story with $J($blue eyes and blond hair$)>$ $J($blue
eyes$)$, a medical example with $J($age over 50 and heart attack$)>J($heart
attack$)$, and a state tax example with $J($increase tax and reduce cigarette
smoking$)>J($reduce cigarette smoking$)$. These results occur using within and
between subject designs; choice, ranking, and rating response methods (Tversky
\& Kahneman, 1983) as well as betting methods \cite{Sides&Osherson_2002}, and
even when participants are paid for being `correct'
\cite{Stolarz-Fantino_2003}. The findings occur with naive (undergraduates)
and sophisticated (e.g. physicians) judges, but it is reduced for participants
who have studied statistics (Tversky \& Kahneman, 1983).

F2. Disjunctive fallacy: $J(H\vee L)<J(H)$ This was found comparing
frequencies \cite{Carlson&Yates_1989} and means \cite{Fisk_2002}. For example,
the Linda story produces $J($feminist or bank teller$)<J($feminist).

F3. Both fallacies together: \ $J(H)>J(H\vee L)>J(H\wedge L)>J(L)$
\cite{Morier&Borgida_1984}. Using the liberal Linda story, Morier and Borgida
(1984) reported the following means: $J($feminist$)=.83$ $>J($feminist or bank
teller$)=.60$ $>J($ feminist and bank teller$)=.36$ $>J($bank teller$)=.26$
($N=64$ observations per mean, and the differences are statistically
significant$)$.

F4. Containment error: $\ J(B)<J(A)$ where $B=(A\vee(\symbol{126}A\wedge B)) $
\cite{Bar-Hillel&Neter_1993}. This was found comparing mean ranks and
frequencies. For example, a photo of an Alpine scene produces $J($photo is
from Switzerland$)>J($photo is from Europe$)$ where of course Europe includes
Switzerland and the rest of Europe other than Switzerland.

F5. Unpacking effects. Implicit subadditivity refers to the order
$J(A)<J((A\wedge B)\vee(A\wedge\symbol{126}B))$. This was found comparing
medians. For example, a story about causes of death produces $J($death by
Homicide$)<J($ death by homicide from an acquaintance $\vee$ death by homicide
from a stranger$)$ \cite{Rottenstreich&Tversky_1997}. The $A$ event is called
the packed event, and the $(A\wedge B)\vee(A\wedge\symbol{126}B)$ event is the
unpacked event.\footnote{This article focuses on implicit
subadditivity/superadditivity, because it only requires an ordinal comparison
of two judgments. Explicit subadditivity/superadditivity
\cite{Tversky-Koehl_Support_2004} is based on the comparison of one judgment
with the sum of several other judgments, and the latter requires much stronger
measurement assumptions. The quantum explanation for implicit unpacking also
applies to explicit unpacking.} However, when the event $A$ is unpacked into
an unlikely event $B$ and a residual, then the opposite effect occurs where
$J(A)>J((A\wedge B)\vee(A\wedge\symbol{126}B))$ \cite{Sloman_Unpack_2004}.

F6. Partitioning effect: The probability judgment given to an event $A$ is
greater when the alternative is described as the negated event $\symbol{126}A
$ (called the case partition) as opposed to a partition equivalent to the
negated event $(B_{1}\vee B_{2}...\vee B_{n})\longleftrightarrow\symbol{126}A$
(called the class partition) \cite{FoxRottenstreich_2003}. This was found
comparing medians. For example, people judge the event `Sunday will be hotter
than any other day next week' (the case based partition) to be greater than
`the hottest day next week will be Sunday' (the class based partition).

F7. Conditional Fallacy: $J(H\mapsto L)<J(L\wedge H).$ For example, when given
a story about an overcast November day in Seattle, the following results were
obtained from 150 participants using medians: $J($it rains)$=.21 $ $>$ $J($ it
rains and temperature remains below $38^{o}F$) $=.18$ $>J( $temperature
remains below $38^{o}F$) $=.14$ $>$ $J($ if it rains then the temperature
remains below $38^{o}F)=.12$ \cite{Miyomoto_condfallacy_1988}. Although the
difference between the medians for the conjunction (.18) and the implication
(.12) is small, it was statistically significant. However, the heart attack
example produces $J($if age is over 50 then heart attack$)=.59>J($age over 50
and heart attack$)=.30>J($heart attack$)=.18$ (Tversky and Kahneman, 1983),
and no differences have also been reported for the example $J($if increase tax
then reduction in cigarette smoking$)=J($increase in tax and reduction in
cigarette smoking$)$ \cite{Hertwig_meaning_of_and_2008}.\ So this result seems
to depend on the type of problem.

F8. Inverse fallacy: $J(A\mapsto B)=J(B\mapsto A)$ . This is found using both
means and frequencies. For example $J($if test is positive then disease is
present$)=J($ if disease is present then test is positive$)$
\cite{Hammerton_1973}. This result occurs with equally likely base rates for
the disease, but unequal likelihoods for the test result given the disease,
and so it is not explained by base rate neglect
\cite{Ville&Mandel_Inverse_2002}. Obviously, this finding also depends on the
type of problem. For example, if a person is murdered, then everyone would
agree that the person is certainly dead; but no one would believe that if a
person is dead, then the person was certainly murdered.

F9. Averaging error: If $J(H)>J(M)>J(L),$ then $J(L)<J(L\wedge M)$ and
$J(H)>J(H\wedge M)$ \cite{Fantino_avg_conj_1997}. This was found comparing
means. For example, using a boring but intellectual Bill story produces
$J($Bill plays in a rock band$)<J($Bill plays in a rock band and Bill is a
park ranger$)$, but $J($Bill builds radio gliders$)>J($ Bill builds radio
gliders and Bill is a park ranger$)$.

F10. Violation of independence: $J(A\wedge C)>J(B\wedge C)$ but $\ J(A\wedge
D)<J(B\wedge D)$ \cite{Miyamoto_1995}. This was found comparing frequencies.
For example, using the story of a\ college applicant named Joe produces
$J($accepted at Harvard and accepted at Princeton$)>J($rejected at Oklahoma
and accepted at Princeton$)$ but $J($accepted at Harvard and rejected at
Texas$)<J($rejected at Oklahoma and rejected at Texas$)$.

F11. Effect of event dependencies. The presence of dependencies between events
$A$ and $B$ affects the rate of conjunction fallacies for $A\wedge B$
\cite{Fisk_2002}. This was found using means and frequencies. A positive
conditional dependency increases the frequency of conjunction errors.

F12. Effect of event likelihoods. a) Highest frequency of conjunction errors
occur with mixed $H\wedge L$ events, a lower frequency occurs with $H\wedge H
$ \ events, and the lowest occurs with $L\wedge L$ events \cite{Wells_1985}.
However, while the mean magnitude of the conjunction error is much larger with
$H\wedge L$ events, no difference is found between $L\wedge L$ and $H\wedge H$
events \cite{Gavanski_1991}. b) The $H\wedge L$ items most often produce only
a single conjunction error with the $L$ event; the $L\wedge L$ event most
often produce zero conjunction errors; and the $H\wedge H$ event produces both
zero and double conjunction errors about equally often
\cite{Yates&Carlson_1986}. But the rate of double conjunction errors with
$H\wedge H$ events is less than 50\%, and they are not found using means
\cite{Gavanski_1991}. The mean estimates for the results reported by Gavansky
and Roskos-Ewoldsen (1991) were $J(A)=.28$, $J(B)=.19$, $J(A\wedge B)=.18$ for
the $L\wedge L$ condition; $J(A)=.77$, $J(B)=.23$, $J(A\wedge B)=.38$ for the
$H\wedge L$ condition; and $J(A)=.76$, $J(B)=.69$, $J(A\wedge B)=.67$ for the
$H\wedge H$ condition. The same general pattern is observed with disjunction
errors -- they are most common and largest in mean magnitude when one event
has a low probability and the other has a high probability \cite{Fisk_2002}.
The mean estimates for the results reported by Fisk (2002) were $J(A)=.36$,
$J(B)=.14\,$, $J(A\vee B)=.27$ for the $L\vee L$ condition; $J(A)=.73$,
$J(B)=.23\,$, $J(A\vee B)=.59$\ for the $H\vee L$ condition; and\ $J(A)=.80$,
$J(B)=.62\,$, $J(A\vee B)=.75$ for the $H\vee H$ condition.

F13. Effect of event relationship. Some researchers find (a) differences
between related and unrelated items (Kahneman \& Tversky, 1983), but (b)
others find a smaller difference (Yates \& Carlson, 1989) or no difference at
all \cite{Gavanski_1991}. An unrelated type of example is to present a boring
Bill story and a liberal Linda story, which produces $J($Bill is an accountant
and Linda is a bank teller$)>J($Linda is a bank teller$)$ as well as $J($Bill
plays jazz and Linda is a feminist$)>J($Bill plays jazz$)$. This was found
using means and frequencies.

F14. Relation to typicality ratings. Conjunction errors correlate with
typicality rating conjunction effects \cite{Shafir&Smith&Osherson_1990}. Same
is true for disjunction errors \cite{Bar-Hillel&Neter_1993}.

F15. Response mode and order effects. Conjunction errors are more prevalent
with ranking than ratings, but there is little or no difference between
probability and frequency ratings \cite{Wedell&Moro_2008}. Apparently the
early finding indicating that frequency formats reduce conjunction errors
confounded class inclusion instructions with ratings versus ranking responses
\cite{Yamagishi_2003}. Conjunction errors are larger in magnitude when the
conjunction is rated first as opposed to being rated last
\cite{Stolarz-Fantino_2003}. This last result can be seen in Figure
\ref{Gavansky_Fig} comparing the circles with the solid dots.

Facts 1 - 9 are considered `errors' with respect to the classic (Kolmogorov)
probability theory. As pointed out by Tversky and Koehler (2004), these facts
seem contrary to other general approaches to judgments of uncertainty
including the theory of belief functions \cite{Shafer_book} as well as fuzzy
set theory \cite{Zadeh_1965}.

\section{Classic Probability Theory}

Before presenting quantum probability theory, it is worth reviewing the basic
assumptions of classic probability theory. This way we can directly compare
the key assumptions underlying the two theories and see exactly where they
differ. A great attraction of classic probabilistic models of cognition is
that they are coherent, that is, predictions are derived from a small set of
axioms \cite{Kolmogorov}. But these models incorporate an important
\textit{hidden} assumption that may be overly restrictive for describing human judgments.

Classic theory provides a set theoretic approach to probabilities: events are
represented as \textit{subsets} from a universal set (called
the\textit{\ sample space}). We will assume that the cardinality of the sample
space is $n$ (a large but finite number). In other words, the sample space
contains $n $ sample points, or unique outcomes (called elements). For this
application, we can think each element, such as $E_{j}$, as representing a
unique pattern of feature values. The story provides information that is used
with prior knowledge to form a \textit{probability function}, denoted
$\mathbf{p}$, which assigns a probability to each element. \ The classic
probability assigned to a particular feature pattern is a positive real number
denoted $p_{j}=\mathbf{p}(E_{j})\geq0$, and these probabilities must sum to
one across all $n$ elements in the universal set. A single question about
event $A$ is represented by a subset, denoted $A^{\prime}$,\textbf{\ }of the
universal event composed of $m\leq n$ elements. The event $A^{\prime} $
contains the subset of elements (feature patterns) that are true for the
question about event $A.$ The classic probability of this event equals the sum
of the elementary probabilities in the subset: $\mathbf{p}(A^{\prime}%
)=\sum_{E_{j}\in A^{\prime}}p_{j}.$ The negation of this event is the set
complement ($\overline{A}$) which has a probability $\mathbf{p}(\overline
{A})=1-\mathbf{p}(A^{\prime})$.

Defining events as sets requires the events to satisfy a set closure property:
If $A^{\prime}$and $B^{\prime}$ are events from the sample space, then the
union and intersections of these two are also events from the sample space.
This brings us to representations for questions about pairs of events. A
question about the conjunction $(A\wedge B)$\ is represented by the
intersection of sets $(A^{\prime}\cap B^{\prime})$, and a question about the
disjunction $(A\vee B)$ is represented by the union of sets $(A^{\prime}\cup
B^{\prime})$. However, this requires making a crucial but hidden assumption
called the \textit{compatibility} assumption. It is assumed that the event
$B^{\prime}$ used for question $B$ is a subset of the \textit{same} sample
space as the subset $A^{\prime}$ used for question $A$. In other words,
different members from a \textit{common} set of elementary events are used to
define $A^{\prime}$ as well as $B^{\prime} $. \ Psychologically, a common set
of features are used to describe both kinds of events. At first it may seem
hard to imagine a situation where compatibility fails, but later we argue that
this key assumption should not be taken for granted. Events defined as sets
satisfy the commutative properties, $(A^{\prime}\cap B^{\prime})=\mathbf{(}%
B^{\prime}\cap A^{\prime})$ and $(A^{\prime}\cup B^{\prime})=(B^{\prime}\cup
A^{\prime}),$ as well as the distributive property $A^{\prime}\cap(B^{\prime
}\cup\overline{B})=(A^{\prime}\cap B^{\prime})\cup(A^{\prime}\cap\overline
{B})$ of Boolean logic.

Conditional probabilities are used to represent judgments about implications
\cite{OberauerWilhelm_2003}. Suppose event $A^{\prime}$ is assumed to be true.
If $A^{\prime}$ is true, then a new conditional probability function
$\mathbf{p}_{A}$ is formed to update the elementary event probabilities as
follows: If $E_{j}\in A^{\prime}$ then $\mathbf{p}_{A}(E_{j})=\mathbf{p}%
(E_{j})/\mathbf{p}(A^{\prime})$ and zero otherwise, so that the sum of the
conditional probabilities equals one. This new conditional probability
function $\mathbf{p}_{A}$\ can then be used to determine new probabilities for
other events from the same sample space. Based on this assumption, the
conditional probability of event $B^{\prime}$ given event $A^{\prime}$ equals
$\mathbf{p}(B^{\prime}|A^{\prime})=\left(  \sum_{E_{j}\in A^{\prime}\cap
B^{\prime}}p_{j}\right)  /\mathbf{p}(A^{\prime})=\mathbf{p}(A^{\prime}\cap
B^{\prime})/\mathbf{p}(A^{\prime}).$

The probability of a \textit{positive} response to the conjunction question
requires yes to question $A$ and a yes to question $B$, which equals the joint
probability $\mathbf{p}(A^{\prime}\cap B^{\prime})=\mathbf{p}(A^{\prime}%
)\cdot\mathbf{p}(B^{\prime}|A^{\prime}).$ A positive response to the
disjunction question requires a yes to $(A\wedge B)$ or $(A\wedge
\symbol{126}B)$ or $(\symbol{126}A\wedge B).$ But a simpler way to answer the
disjunction is to make a \textit{negative} response if the answer to question
$A$ is no and the answer to question $B$ is no, so that $\mathbf{p}(A^{\prime
}\cup B^{\prime})=1-\mathbf{p}(\overline{A}\cap\overline{B}).$ The latter is
particularly useful when more events are involved and so we will use this form
hereafter.\ The \textit{law of total} probability, which is a key principle
for Bayesian modeling, follows from the distributive law of Boolean logic:
\begin{align}
\mathbf{p}(B^{\prime})  & =\mathbf{p}(B^{\prime}\cap(A^{\prime}\cup
\overline{A}))=\mathbf{p}((B^{\prime}\cap A^{\prime})\cup(B^{\prime}%
\cap\overline{A}))\label{Law of Total Prob}\\
& =\mathbf{p}(A^{\prime}\cap B^{\prime})+\mathbf{p}(\overline{A}\cap
B^{\prime})=\mathbf{p}(A^{\prime})\cdot\mathbf{p}(B^{\prime}|A^{\prime
})+\mathbf{p}(\overline{A})\cdot\mathbf{p}(B^{\prime}|\overline{A}).\nonumber
\end{align}
The above probability rules imply the following orders: $1\geq\mathbf{p}%
(H^{\prime}\cup L^{\prime})\geq\mathbf{p}(H^{\prime})\geq\mathbf{p}(L^{\prime
})\geq\mathbf{p}(H^{\prime}\cap L^{\prime})\geq0$, and $\mathbf{p}(A^{\prime
}|B^{\prime})\geq\mathbf{p}(A^{\prime}\cap B^{\prime})=\mathbf{p}(B^{\prime
})\cdot\mathbf{p}(A^{\prime}|B^{\prime})$.

The prime notation $A^{\prime}$ was introduced by Tversky and Koehler (1994)
to distinguish questions about an event $A$ from the corresponding
mathematical set $A^{\prime}$ implied by the description. This is needed
because two different descriptions could logically imply the same set, yet
judgments may differ between the two logically equivalent descriptions. For
similar reasons, different symbols are used to represent conjunctive and
disjunctive questions $(\wedge,\vee)$ and the corresponding intersection and
union relations $(\cap,\cup)$ used in classic probability theory. This is
necessary because the logical relations implied by these symbols may obey
different observable properties. If we assume natural language conjunction
$(A\wedge B)$ corresponds with intersection $(A^{\prime}\cap B^{\prime})$ and
natural language disjunction $(A\vee B)$ corresponds with union $(A^{\prime
}\cup B^{\prime})$, then facts 1-9 show that human judgments do not follow
classic probability theory. One way to retain a classic probability theory of
human judgment in view of these facts is to assume that such direct and strict
correspondences do not hold \cite{Krynski&Tenenbaum_2007}. For example, one
can assume that the conjunction question is answered using a conditional
probability of the story given the event in question \cite{Fisk_1996}. In
other words, people misinterpret the questions and judge the wrong
probabilities. But this argument does not apply to studies that use betting
procedures, which implicitly require likelihoods to make decisions, and never
explicitly request a probability judgment. Another way to retain classic
probability theory is to assume that each single probability judgment from an
individual follows classic rules, but these judgments are based on noisy
sample estimates contaminated by error \cite{ErevWallstenBudescu_1994}. Noisy
probability estimates can produce highly frequent conjunction errors
\cite{Costello_2009}. However, this cannot explain violations of conjunction
and disjunction rules when these violations occur with means and medians which
cancel out the noise.

\section{Quantum Probability Theory}

First we will briefly summarize the basic assumptions of quantum probability
theory. This summary has to be abstract so that we can compare only the
essential and basic assumptions directly with classic probability theory.
Later we elaborate with simple graphical and numerical examples and provide
important psychological intuitions behind these ideas.\footnote{The reader
only needs knowledge of linear algebra to understand this section. We realize
that some readers may need reminders and so we included a brief tutorial in
the appendix. No knowledge of physics is required. This application only uses
the simplest and most basic ideas of quantum theory. See Hughes for a good non
physical introduction to quantum theory \cite{Hughes_1989}.} Quantum theory is
comparable with classic probability theory in terms of it's coherence -- it's
predictions are also derived from a small set of axioms \cite{VonNeumann_1932}%
. \ But quantum axioms differ from classic axioms, and it is an empirical
question whether one or the other provides a better representation of human judgment.

Quantum theory provides a geometric approach to probabilities: events are
represented by \textit{subspaces} of a \textit{vector space} (called the
Hilbert space). We will assume that the dimensionality of the vector space is
$n$ (again a large but finite number). In other words, the vector space is
based on $n$ orthogonal and unit length vectors (called eigenvectors). For
this application, we can think of each eigenvector, denoted $\mathbf{V}_{j}$,
as representing a unique pattern of feature values.\ The story provides
information that is used with prior knowledge to form a \textit{state vector},
denoted $\mathbf{\psi}\,$, which assigns a scalar (called an amplitude) to
each eigenvector by the inner product $\mathbf{V}_{j}^{\dagger}\cdot
\mathbf{\psi}=$ $\psi_{j}$. The quantum probability of a particular feature
pattern equals the squared magnitude of its amplitude, $q(\mathbf{V}_{j})=$
$\left\vert \psi_{j}\right\vert ^{2}$, and these probabilities must sum to one
across all $n$ eigenvectors of the vector space (this is called Born's rule).
A single question about event $A$ is represented by an $m$-dimensional
subspace, denoted $A"$, within the vector space ($m\leq n$). The subspace $A"$
is spanned by a subset of the eigenvectors (feature patterns) that are true
for the question about the event $A$. \ The quantum probability for this event
equals the sum of the squared magnitudes of the amplitudes for the
eigenvectors that span the subspace: $q(A")=\sum\nolimits_{V_{j}\in
A"}\left\vert \psi_{j}\right\vert ^{2}$. The negation of this event is the
$(n-m)$ dimensional subspace, denoted $A^{\perp}$, that is orthogonal to the
subspace $A",$ which has a probability $q(A^{\perp})=1-q(A").$

Defining events as subspaces implies that the events must satisfy a subspace
closure property: if vectors $\mathbf{V}_{j}$ and $\mathbf{V}_{k}\,\ $are
members of the subspace, then $\mathbf{W}=a\mathbf{V}_{j}+b\mathbf{V}_{k}$,
for arbitrary scalars $a,b$ must also be a member. Consequently, one full set
of eigenvectors $\{\mathbf{V}_{j}$, $j=1,n\}$ can be `rotated' by a unitary
(orthonormal) matrix into another full set of eigenvectors $\{\mathbf{W}_{j},$
$j=1,n\}$. Thus there exists more than one set of eigenvectors that can be
used to describe events within the \textit{same} vector space. This brings us
again to representations of questions about pairs of events. Suppose question
$A$ corresponds to subspace $A"$ described by a subset of the $\mathbf{V}_{j}$
eigenvectors; but suppose question $B$ corresponds to a subspace $B"$ that
cannot be described by these same features, and instead it requires a
different subset of the $\mathbf{W}_{j}$ eigenvectors. Then the pair of events
$A"$, $B"$ cannot be described by a common set of eigenvectors, which makes
these two events \textit{incompatible}. Psychologically, different kinds of
features may be needed to describe the two different events. If event $A"$ can
be defined by the same set of $\mathbf{V}_{j}$ eigenvectors as event $B"$,
that is they share a common set of eigenvectors, then these two events are
\textit{compatible}. Quantum theory requires a general representation of
conjunction and disjunction that applies to both compatible and incompatible
events. This is achieved by using a sequential logical operation to represent
conjunction and disjunction questions \cite{Franco_JMP_2009}.\footnote{One
might wonder if it makes sense to represent conjunction by the span of the
intersection of two sets of eigenvectors, and to represent disjunction by the
span of the union of two sets of eigenvectors. There are two major objections
for incompatible events. First, according to Bohr's principle of
complementarity, incompatible events cannot be evaluated simultaneously, and
they must be examined sequentially. Second, this fails empirically to explain
the conjunction and disjunction fallacies.} Suppose a question about event $A$
is asked first followed by a question about event $B$. These questions are
answered in order and the requested logical operation is performed on the
answers. If asked about the conjunction in this order, then a
\textit{positive} response to the conjunction requires a yes to $A$ followed
by a yes to $B$, and this sequential logical and operation is denoted
$(A"\sqcap B")$. If asked about the disjunction in this order, then a
\textit{negative} response to the disjunction requires a no to $A$ followed by
a no to $B$. A \textit{positive} response to the logical disjunction in this
order is denoted $(A"\sqcup B").$ If the events are compatible, then the
commutative property holds $(A"\sqcap B")=(B"\sqcap A")$ and $(A"\sqcup
B")=(B"\sqcup A")$, and so does the distributive property $A"\sqcap(B"\sqcup
B^{\perp})=(A"\sqcap B")\sqcup(A"\sqcap B^{\perp})$ (see Appendix). But if the
events are incompatible, then both of these properties fail. Therefore,
quantum events only obey a partial Boolean algebra \cite{Hughes_1989}.

Conditional quantum probabilities are used to represent judgments about
implications. Suppose event $A"$ is assumed to be true, which is defined in
terms of the $\mathbf{V}_{j}$ eigenvectors. If $A"$ is true, then a new
conditional state vector $\mathbf{\psi}_{A}$ is formed which is defined as
follows: If $\mathbf{V}_{j}\in A"$ then the new amplitude assigned to
$\mathbf{V}_{j\text{ }}$equals $\mathbf{V}_{j}^{\dagger}\cdot\mathbf{\psi}%
_{A}=$ $\psi_{j}/\sqrt{q(A")}$ and zero otherwise, so that the sum of the
conditional probabilities equals one (von Neumann called this state
reduction). Now suppose we want to determine the probability of a new event
$B",$ which is defined by the $\mathbf{W}_{j}$ eigenvectors. Then the
probabilities for the new event $B"$ given $A"$ equals $q(B"|A")=\sum
\nolimits_{W_{i}\in B"}\left\vert \mathbf{W}_{i}^{\dagger}\cdot\mathbf{\psi
}_{A}\right\vert ^{2}$ (called L\"{u}der's rule).

The probability of a positive response to a conjunction equals the probability
of saying yes to the sequence of questions, $q(A"\sqcap B")=q(A")\cdot
q(B"|A")$. The probability of a negative response to a disjunction equals
$q(A^{\perp}\sqcap B^{\perp})=q(A^{\perp})\cdot q(B^{\perp}|A^{\perp})$ and so
the probability of a positive response to the disjunction is $q(A"\sqcup
B")=1-q(A^{\perp}\sqcap B^{\perp}).$ If the events are compatible, then
quantum probability obeys the same laws as classic probability (see Appendix),
but if the events are incompatible they do not (see the examples below). \ 

In summary, the two probabilities theories share many similarities. Both
provide principles for defining probabilities for single events, complements,
conjunctions, disjunctions, and implications (conditional probabilities).
However, the key differences are that classic probability represents events as
sets, which forces all the events to be compatible so that they satisfy the
commutative and distributive properties of Boolean algebra; whereas quantum
theory represents events as subspaces, which allows events to be either
compatible or incompatible, and the latter can violate the commutative and
distributive properties of Boolean algebra. But this has been presented in a
very abstract manner to compare basic assumptions, and next we give a more
intuitive presentation of quantum theory.

\subsection{Simple illustration of quantum principles}

\begin{figure}
[ptb]
\begin{center}
\includegraphics[
natheight=4.968400in,
natwidth=6.562200in,
height=5.0237in,
width=6.6262in
]%
{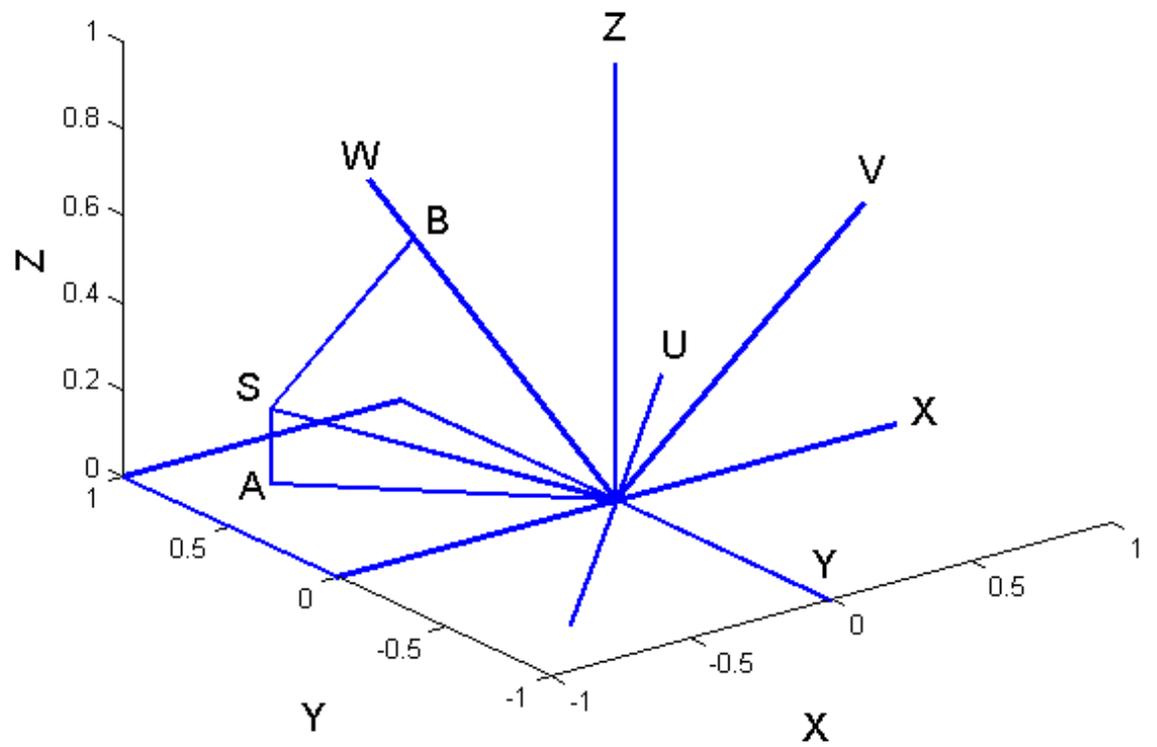}%
\caption{Three dimensional vector space with two sets of incompatible
questions}%
\label{projection}%
\end{center}
\end{figure}

Figure \ref{projection} provides a simple illustration of all the ideas using
only a three dimensional vector space. (In general, we do not necessarily
assume such a simple space). It is most convenient to use the matrix algebra
of projectors to do quantum calculations (using Matlab, or R, or Mathematica, ect.).%

\qquad

The set of three orthogonal axes labeled $\{\mathbf{X,Y,Z}\}$ represent three
different eigenvectors. For example, these three eigenvectors could represent
three mutually exclusive and exhaustive responses for a voter such as
democrat, republican, or independent (for our purposes, independent means not
democrat or republican):%

\[
\mathbf{X}=%
\begin{bmatrix}
1\\
0\\
0
\end{bmatrix}
,\mathbf{Y}=%
\begin{bmatrix}
0\\
1\\
0
\end{bmatrix}
,\mathbf{Z}=%
\begin{bmatrix}
0\\
0\\
1
\end{bmatrix}
.
\]
The quantum state $\mathbf{\psi}$ is represented in this case by the vector
associated with the letter $\mathbf{S}$ in the figure (e.g., the state of an
undecided voter just before a presidential election). This state can be
described in terms of coordinates with respect to the $\{\mathbf{X,Y,Z\}}$
eigenvectors. In this case, the state assigns the following amplitudes to the
$\{\mathbf{X,Y,Z\}}$ eigenvectors%
\begin{equation}
\mathbf{\psi}=\mathbf{S}=(-.6963)\cdot\mathbf{X}+(.6963)\cdot\mathbf{Y}%
+(.1741)\cdot\mathbf{Z}.\label{SuperpositionState1}%
\end{equation}
Dirac called vectors, such as Equation \ref{SuperpositionState1},
\textit{superposition} states with respect to the eigenvectors
$\{\mathbf{X,Y,Z\}}$. At this point, a superposition state simply assigns
probabilities to events generated by $\{\mathbf{X,Y,Z\}}$, but later this
concept takes on a deeper meaning. As can be seen in this example, the
amplitudes assigned to each eigenvector can be positive or negative (or even
complex numbers). But note that the length of the state vector $\mathbf{S}$
equals one.

Quantum probabilities are expressed most simply and intuitively by using the
geometric concept of a projection (see Appendix). In this example, the state
vector $\mathbf{S}$ lies in a three dimensional space. The events of interest
are subspaces that have smaller dimension (rays or planes in this case). To
determine the probability of an event, we first project the state vector on to
the subspace that represents the event and then compute its squared length. A
matrix is used to perform this projection, which is called the
\textit{projector} for the event. All the quantum events generated by the
$\{\mathbf{X,Y,Z}\}$ eigenvectors are based on the following three projectors
(formed by outer products):%

\begin{gather*}
M_{X}=\mathbf{X}\cdot\mathbf{X}^{\dagger},\ M_{Y}=\mathbf{Y}\cdot
\mathbf{Y}^{\dagger},\ M_{Z}=\mathbf{Z}\cdot\mathbf{Z}^{\dagger}\\
M_{X}+M_{Y}+M_{Z}=I\ (identity).
\end{gather*}
For example, $M_{X}$ is a $3\times3$ projector matrix (it has a one in the
first row and column, and zeros everywhere else), and it projects the state
$\mathbf{S}$ on to the ray $X"$ containing the $\mathbf{X}$ eigenvector. The
projection equals the matrix product of projector and the state, $M_{X}%
\cdot\mathbf{S.}$ The probability of event $X"$ (e.g., democrat) equals the
square length of the projection $\left\vert M_{X}\cdot\mathbf{S}\right\vert
^{2}=|-.6963|^{2}=.4848$. The probability of event $Y"$(e.g. republican) is
computed in the same way, $\left\vert M_{Y}\cdot\mathbf{S}\right\vert ^{2}$
$=|.6963|^{2}$ $=.4848$ . Suppose the quantum event in question is the plane
formed by the span of eigenvectors $\{\mathbf{X},\mathbf{Y\}}$, which is
symbolized as $X"+Y"$ (e.g., `democrat or republican'). The projection of the
state $\mathbf{S}$ onto the $X"+Y"$ plane is the vector associated with the
label $\mathbf{A}$. \ The matrix $M_{X+Y}=M_{X}+M_{Y}$ projects the state
vector $\mathbf{S}$\textbf{\ }onto the subspace $X"+Y"$:%

\[
\mathbf{A}=M_{X+Y}\cdot\mathbf{S}=%
\begin{bmatrix}
1 & 0 & 0\\
0 & 1 & 0\\
0 & 0 & 0
\end{bmatrix}
\cdot\mathbf{S}=%
\begin{bmatrix}
-.6963\\
.6963\\
0
\end{bmatrix}
,
\]
and the probability of the $X"+Y"$ event equals $|\mathbf{A}|^{2}=$
$|-.6963|^{2}+\left\vert .6963\right\vert ^{2}=.9697$. The negation of this
disjunction event is the ray associated with the $\mathbf{Z}$ eigenvector
(e.g., independent), and the probability of this event is $|M_{(X+Y)^{\perp}%
}\mathbf{\psi}|^{2}=\left\vert (I-M_{X+Y})\cdot\mathbf{S}\right\vert
^{2}=|M_{Z}\cdot\mathbf{S}|^{2}=|.1741|^{2}=.0303=1-.9697$.

If we were restricted to use only projections on the $\{\mathbf{X,Y,Z}\}$
eigenvectors, then quantum probabilities would obey the same laws as classic
probabilities. However, a vector space has \textit{no} privileged set of
eigenvectors. We could rotate the first set $\{\mathbf{X,Y,Z}\}$ of
eigenvectors to form a new orthonormal set of eigenvectors labeled
$\{\mathbf{U,}$ $\mathbf{V,W}\}$ in the figure. The unitary transformation
matrix that generates coordinates for the new eigenvectors $\{\mathbf{U,}$
$\mathbf{V,W}\}$ is%

\[
T=%
\begin{bmatrix}
1/\sqrt{2} & 1/2 & -1/2\\
1/\sqrt{2} & -1/2 & 1/2\\
0 & 1/\sqrt{2} & 1/\sqrt{2}%
\end{bmatrix}
=%
\begin{bmatrix}
\mathbf{U} & \mathbf{V} & \mathbf{W}%
\end{bmatrix}
.
\]
The first column of $T$ gives the coordinates of the $\mathbf{U}%
=T\cdot\mathbf{X}$ eigenvector, which is the ray that runs through the main
diagonal of the $X"+Y"$ plane. The second and third columns of $T$ give the
$\mathbf{V}=T\cdot\mathbf{Y}$ and $\mathbf{W}=T\cdot\mathbf{Z}$ eigenvectors.
This new set of eigenvectors $\{\mathbf{U,V,W}\}$ represents a different
perspective for understanding features (e.g., moderate, liberal, conservative
). In this (artificial) example, the eigenvector $\mathbf{U}$ (e.g. moderate)
lies in the $X"+Y"$ plane and happens to be midway between the eigenvectors
$\mathbf{X}$ and $\mathbf{Y}$ (e.g. democrat, republican). All the quantum
events generated from the $\{\mathbf{U,V,W}\}$ \ set of eigenvectors are based
on the following three projectors (again formed by outer products):%
\begin{gather*}
M_{U}=\mathbf{U}\cdot\mathbf{U}^{\dagger},\ M_{V}=\mathbf{V}\cdot
\mathbf{V}^{\dagger},\ M_{W}=\mathbf{W}\cdot\mathbf{W}^{\dagger}\\
M_{U}+M_{V}+M_{W}=I.
\end{gather*}
The state vector $\mathbf{\psi}=\mathbf{S}$ can also be described in terms of
the amplitudes assigned to the new eigenvectors $\{\mathbf{U,V,W}\}.$ The
matrix product
\[
T^{\dagger}\cdot\mathbf{S=}%
\begin{bmatrix}
\mathbf{U}^{\dagger}\\
\mathbf{V}^{\dagger}\\
\mathbf{W}^{\dagger}%
\end{bmatrix}
\cdot\mathbf{S=}%
\begin{bmatrix}
\mathbf{U}^{\dagger}\cdot\mathbf{S}\\
\mathbf{V}^{\dagger}\cdot\mathbf{S}\\
\mathbf{W}^{\dagger}\cdot\mathbf{S}%
\end{bmatrix}
=%
\begin{bmatrix}
0\\
-.5732\\
.8194
\end{bmatrix}
.
\]
transforms the amplitudes originally assigned to eigenvectors $\{\mathbf{X}%
,\mathbf{Y},\mathbf{Z}\}$ into amplitudes assigned to eigenvectors
$\{\mathbf{U},\mathbf{V},\mathbf{W}\}$. This allows us to express the state
$\mathbf{S}$ as a superposition state with respect to the eigenvectors
$\{\mathbf{U,V,W\}} $. In other words, the exact same state $\mathbf{S}$ can
be expressed as a superposition of the $\{\mathbf{X,Y,Z\}}$ eigenvectors or as
a superposition of the $\{\mathbf{U,V,W\}}$ eigenvectors:%
\begin{gather}
(-.6963)\cdot\mathbf{X}+(.6963)\cdot\mathbf{Y}+(.1741)\cdot\mathbf{Z=S}%
\label{SuperpositionState2}\\
=I\cdot\mathbf{S}=(M_{U}+M_{V}+M_{W})\cdot\mathbf{S}\nonumber\\
=\mathbf{U}\cdot\mathbf{U}^{\dagger}\cdot\mathbf{S}+\mathbf{V}\cdot
\mathbf{V}^{\dagger}\cdot\mathbf{S}+\mathbf{W}\cdot\mathbf{W}^{\dagger}%
\cdot\mathbf{S}\nonumber\\
=0\cdot\mathbf{U}+(-.5732)\cdot\mathbf{V}+(.8194)\cdot\mathbf{W.}\nonumber
\end{gather}
Now the concept of superposition becomes much deeper because the same state
$\mathbf{S}$ must generate probabilities for \textit{two} different sets of
eigenvectors. Shortly we will show how superposition states with respect to
\textit{different} sets of eigenvectors produce interference effects that are
critical for explaining violations of classic probability theory. But first,
let us continue with a few more example calculations. The probability of the
event $W"$ (e.g. conservative) is determined by projecting the state
$\mathbf{S}$ on to the eigenvector $\mathbf{W}$, using the projector $M_{W}$,
which produces the projection associated with the vector labeled $\mathbf{B}$
in the figure. The squared length of this projection equals
\begin{align*}
q(W")  & =|\mathbf{B}|^{2}=|M_{W}\cdot\mathbf{S}|^{2}\\
& =|\mathbf{W}\cdot\mathbf{W}^{\dagger}\cdot\mathbf{S}|^{2}\\
& =|\mathbf{W}|^{2}\cdot|\mathbf{W}^{\dagger}\cdot\mathbf{S}|^{2}%
=1\cdot|\mathbf{W}^{\dagger}\cdot\mathbf{S}|^{2}\\
& =|.8194|^{2}=.6714,
\end{align*}
which is simply the squared magnitude of the amplitude assigned to
$\mathbf{W}$ in Equation \ref{SuperpositionState2}.

Finally consider the conditional probability of event $W"$ given that event
$X"+Y"$ has occurred (e.g., the probability that the voter is conservative
given that the person voted for a democrat or republican). Following
L\"{u}der's rule, we first compute the normalized projection of the original
state $\mathbf{S}$ on the known event $X"+Y"$. Recall from above that
$\mathbf{A}$ is the projection, and this is normalized to form the vector
$\mathbf{\psi}_{X+Y}=\mathbf{A}/\left\vert \mathbf{A}\right\vert =%
\begin{bmatrix}
-1/\sqrt{2} & 1/\sqrt{2} & 0
\end{bmatrix}
^{\dagger}.$ Then we compute the squared length of the projection of the state
$\mathbf{\psi}_{X+Y}$ onto the ray $W"$, which equals $q(W"|X"+Y")=\left\vert
M_{W}\cdot\mathbf{\psi}_{X+Y}\right\vert ^{2}$ $=|(W\cdot W^{\dagger}%
\cdot\mathbf{A}/|\mathbf{A}|)|^{2}=.50$. \ Similarly, the probability
$q(W"|Z")=$ $\left\vert M_{W}\cdot\mathbf{Z}\right\vert ^{2}=.50.$

A \textit{cognitive} \textit{processing} interpretation of the basic quantum
principles can now be given by using the geometric concepts of states and
projectors. The eigenvectors correspond to feature patterns that are used to
describe or characterize events. The initial state vector $\mathbf{\psi}$
represents the memory trace that determines the potential for a pattern to be
retrieved, which is formed by the person's prior knowledge and the story told
to the person. When questioned about a single event $A$, a projector
$M_{A}=\sum\mathbf{V}_{j}\cdot\mathbf{V}_{j}^{\dagger}$ is formed from the
features (eigenvectors) $\{\mathbf{V}_{j}$, $j\in A"\}$ representing the
question $A$. The projection, $M_{A}\cdot\mathbf{\psi}$, determines how well
the retrieval cue provided by the question matches the memory state, and the
probability of a retrieval equals the squared length of the projection:
$q(A")=\left\vert M_{A}\cdot\mathbf{\psi}\right\vert ^{2}=\sum_{V_{j}\in
A"}\left\vert \psi_{j}\right\vert ^{2}$. If event $A"$ is assumed to be true,
then the initial state $\mathbf{\psi}$ changes to a new state $\mathbf{\psi
}_{A}$, which is the normalized projection $\mathbf{\psi}_{A}=$ $(M_{A}%
\cdot\mathbf{\psi})/\left\vert M_{A}\cdot\mathbf{\psi}\right\vert .$ After
given this information, if a second question is asked about event $B$, then a
projector $M_{B}=\sum\mathbf{W}_{j}\cdot\mathbf{W}_{j}^{\dagger}$ is formed
from the features (eigenvectors) $\{\mathbf{W}_{j} $, $j\in B"\}$ representing
the question $B$, and the conditional probability of a positive response to
the retrieval cue $B"$ after given information about $A"$ equals
$q(B"|A")=\left\vert M_{B}\cdot\mathbf{\psi}_{A}\right\vert ^{2}$. Finally,
the probability of a positive response to the conjunction equals the
probability of positive retrievals to both the first and second questions
\begin{align}
q(A"\sqcap B")  & =q(A")\cdot q(B"|A")\label{QPand}\\
& =\left\vert M_{A}\cdot\mathbf{\psi}\right\vert ^{2}\cdot\left\vert
M_{B}\cdot\mathbf{\psi}_{A}\right\vert ^{2}\nonumber\\
& =\left\vert M_{A}\cdot\mathbf{\psi}\right\vert ^{2}\cdot\left\vert
M_{B}\cdot\frac{(M_{A}\cdot\mathbf{\psi})}{\left\vert M_{A}\cdot\mathbf{\psi
}\right\vert }\right\vert ^{2}\\
& =\left\vert M_{B}\cdot M_{A}\cdot\mathbf{\psi}\right\vert ^{2}.\nonumber
\end{align}

\subsection{Interference}

The possibility of using different sets of eigenvectors, $\{\mathbf{X,Y,Z}\}$
versus $\{\mathbf{U}$ ,$\mathbf{V}$,$\mathbf{W}\}$, within the same vector
space to represent different types of questions introduces an important
psychological issue about the disturbance or interference of one question by
another. Suppose we ask about a question about event $W$ (e.g. conservative).
Before we ask this question, while the person is in the initial state
$\mathbf{S}$, there is a .6714 probability of answering yes. After obtaining
this answer, the state changes (according to L\"{u}der's rule) from state
$\mathbf{S}$ to state $\mathbf{W}$. If we ask the same question again
immediately, the person will answer yes with certainty (and with frustration
for being asked to repeat the answer). The state $\mathbf{W} $ is no longer in
superposition with respect to the $\{\mathbf{U,V,W\}}$ eigenvectors. However,
this same state $\mathbf{W}$ is in superposition with respect to the
$\{\mathbf{X,Y,Z\}}$ eigenvectors. In other words, once a person becomes
certain about the $\{\mathbf{U,V,W\}}$ set of eigenvectors, this person
\textit{must} become uncertain with respect to the $\{\mathbf{X,Y,Z\}}$ set of
eigenvectors. The person can't be certain about both at the same time
(Heisenberg called this the uncertainty principle). Furthermore, if we now ask
a question about event $Y$ (e.g., republican), then the probability of a yes
equals .25 (given state $\mathbf{W}$). If the answer to the $Y$ question
happens to be yes, then the state changes (according to L\"{u}der's rule
again) from $\mathbf{W}$ to $\mathbf{Y},$ and the person becomes certain about
the $Y$ question, but becomes uncertain again about the $W$ question
(probability of yes to question $W$ given state $\mathbf{Y}$ is also .25). In
other words, asking the question about $Y$ after the answer to question $W$
has changed the likelihood of responses about question $W$ from certainty to
uncertainty again. (This is the reason filler items are inserted in between
repetitions of a question). This type of disturbance between questions can
always happen with superposition states described by different sets of
eigenvectors within the same vector space.

\subsection{Compatibility of events}

According to quantum theory, order is usually critical, and one has to be
careful of the order that questions are asked. For example, a projection on
$X"$ followed by a projection on $U"$is not the same as these operations in
reverse ($M_{X}\cdot M_{U}\neq M_{U}\cdot M_{X}$, i.e., the projection
matrices do not commute). In other words, asking question $X$ first (e.g.
whether a person is a democrat or not) followed by asking question $U$ (e.g.,
whether a person is a moderate or not) is not necessarily the same as asking
these questions in the opposite order. This order effect indicates the
property of \textit{incompatibility} between these two events (they do not
share the same eigenvectors). Psychologically, one can only view \textit{one}
perspective at a time, the questions \textit{must} be answered sequentially,
and as we have seen, asking one question from one perspective can
\textit{disturb} a later question from a different perspective (for example,
first asking about being a moderate can disturb a later question about being a
democrat). There is an abundance of research demonstrating order effects on
probability judgments \cite{Hogarth&Einhorn_1992}. For example, when judging
probabilities of guilt in a criminal trial, the direction of the effect of
weak evidence on judgments depends on whether it precedes or follows strong
evidence \cite{McKenzieLeeChen_2002}. These order effects are inconsistent
with classic probability theory, and in the past, they have been explained in
terms of anchoring and adjustment type of adding or averaging models.

This capability of changing eigenvectors (i.e. changing perspectives) and
producing incompatible events makes quantum theory fundamentally different
than classic theory. Classic (Kolmogorov) probability theory assumes a single
compatible representation of events. Figure \ref{projection} was constructed
assuming that quantum events $X",Y",Z"$ (e.g., democrat, republican,
independent) are incompatible with the quantum events $U",V",W"$(e.g.,
moderate, liberal, conservative). In this case, one set of events is a
rotation of the other set of events within the same three dimensional space as
depicted in Figure \ref{projection}. To evaluate a question about the $X$
event, we need to adopt the $\{\mathbf{X,Y,Z}\}$ eigenvector point of view;
but then to evaluate a question about the $U$ event, we need to rotate to the
$\{\mathbf{U,V,W}\}$ eigenvector perspective. One cannot evaluate questions
about $X$ and $U$ simultaneously (Bohr called this the principle of complementarity).

Alternatively, note that whenever we ask questions using eigenvectors from the
same set, then the order does not matter. For example, $M_{X}\cdot M_{Y}%
=M_{Y}\cdot M_{X}$ so $X"$ (e.g. is a person a democrat) is compatible with
$Y"$ (e.g., is a person a republican). This lack of order effect defines the
property of \textit{compatibility} between these two events (they share the
same eigenvectors). In this case one can maintain the same perspective while
answering both questions. In this way, the questions can be answered
simultaneously, because one question does not disturb the other. In this case
the two events $X",Y"$ are also mutually exclusive (i.e., orthogonal
subspaces), and so are the two events $U",V"$. In general, if two events
$M_{A},M_{B}$ are orthogonal to each other, then they are compatible because
$M_{A}\cdot M_{B}=0=M_{B}\cdot M_{A}.$ However, it is possible that two events
can be compatible yet not orthogonal. For example, $M_{U+V}\cdot M_{V+W}%
=M_{V}=M_{V+W}\cdot M_{U+V}$ \ and so $U"+V"$ (e.g., is a person a moderate or
liberal) is compatible with $V"+W"$ (e.g., a person is liberal or conservative).

Now let us turn and examine the geometric situation used to represent events
when they are \textit{all} compatible. Once again suppose that $\{X",Y",Z"\}$
represent three mutually exclusive and exhaustive quantum events (e.g., a
person is a democrat, republican, independent); and suppose that
$\{Q",R",S"\}$ is a different set of three mutually exclusive and exhaustive
quantum events (e.g., a person is young, middle age, or old). As before, the
events in $\{X",Y",Z"\}$ are not necessarily orthogonal to the events in
$\{Q",R",S"\}$, \ but now we assume that the events in $\{X",Y",Z"\}$ are
compatible with the events in $\{Q",R",S"\}.$ This implies not only that
$M_{X}\cdot M_{Y}=M_{Y}\cdot M_{X}$ but also that $M_{X}\cdot M_{Q}=M_{Q}\cdot
M_{X}$, and this is true for all pairs of events. Now it is
\textit{impossible} to represent all these events by Figure \ref{projection},
because all of these compatible properties cannot occur within a 3 dimensional
space. These compatible events require (at least) a 9-dimensional vector space
(see Appendix) based on 9 orthonormal eigenvectors $\{\mathbf{XQ}%
,\mathbf{XR},\mathbf{XS}$\textbf{,}$\mathbf{YQ},\mathbf{YR},\mathbf{YS}%
$\textbf{,}$\mathbf{ZQ},\mathbf{ZR},\mathbf{ZS}\}$, which is forms a
\textit{tensor} product space. In this 9-dimensional vector space, the single
ray or eigenvector $\mathbf{XQ}$ represents the pattern or joint event $X"\cap
Q"$ (e.g. democrat and young), and the amplitude assigned to the $\mathbf{XQ}$
eigenvector determines the joint probability of $X"\cap Q"$. Using this
representation, the event $X"=XQ"+XR"+XS"$ (e.g., democrat) corresponds to the
projector $M_{X}=(M_{XQ}+M_{XR}+M_{XS})$, the event $Q"=XQ"+YQ"+ZQ"$ (e.g.,
young) corresponds to the projector $M_{Q}=(M_{XQ}+M_{YQ}+M_{ZQ})$,\ the
intersection event $(X"\cap Q")=XQ"$ corresponds to the projector $M_{X\cap
Q}=M_{X}\cdot M_{Q}=$ $M_{Q}\cdot M_{X}=M_{XQ}$ $=\mathbf{XQ}\cdot
\mathbf{XQ}^{\dagger},$ and the span $X"+Q" $ corresponds to $M_{X+Q}%
=M_{X}+M_{Q}-M_{Q}\cdot M_{X}$. If all of the events are compatible, then the
probabilities computed from quantum theory obey the same laws as the
probabilities computed from classic (Kolmogorov) theory (see Appendix).

In summary, if events $\{X",Y",Z"\}$ and events $\{U",V",W"\}$ are
incompatible, then the person can only respond with one of three possible
outcomes at any point in time. The person can choose a response from the set
$\{X",Y",Z"\},$ or the person can choose a response from the set
$\{U",V",W"\},$ but we cannot observe any combinations. So this situation can
be represented within a 3-dimensional space. But when the events in
$\{X",Y",Z"\}$ and $\{Q",R",S"\}$ are compatible, then a person can respond
with a pair, one from each set, which means one of 9 possible outcomes can
occur. So we need to use at least a 9 - dimensional space to represent this
situation. These are the smallest possible dimensions that could be used for
these examples, and in general, the dimensionality could be much larger in
both cases.

Of course it is possible to have a combination of compatible and incompatible
events. For example, suppose we had three sets of questions: a first set of
mutually exclusive and exhaustive events $\{X",Y",Z"\}$, a second set of
mutually exclusive and exhaustive events $\{U",V",W"\}$, and a third set of
mutually exclusive and exhaustive events $\{Q",R",S"\}$. Again we suppose that
a question taken from one set is not orthogonal to a question taken from a
different set. In this situation it is possible, for example, to have the
first and second set be incompatible with each other, but both could be
compatible with the third set. This situation would require at least a
9-dimensional space. This vector space would be spanned by 9 eigenvectors
formed from combinations of the first and third sets, or it would be spanned
by 9 eigenvectors formed from combinations of the second and third sets;
furthermore the two sets of eigenvectors would be related by a unitary transformation.

When should events be treated as compatible or incompatible? The general
answer is that this is an empirical question, and order effects are an
empirical sign of incompatibility. However, at this point we make the working
hypothesis that compatibility depends on experience with the combination of
events. Conjunction errors disappear when individuals are given direct
training experience with pairs of events \cite{Nilsson_2008}, and order
effects on abductive inference also decrease with training experience
\cite{WangToddZhang_2006}. On the one hand, if the person has a great deal of
experience with the combination or pattern of events, then they have the
opportunity to form a compatible vector space, and they can estimate the
intersection of events from this large space of patterns of events. On the
other hand, if an unusual or novel combination of events is presented, and the
person has little or no experience with such combinations, then they may not
have formed a compatible representation, and they must rely on incompatible
representations of events that use the same small vector space but require
taking different perspectives. A second way to facilitate the formation of a
compatible representation is to present the required joint frequency
information in a tabular format \cite{WolfeReyna_2009}. Instructions to use a
joint frequency table format would encourage a person to form and make use of
a compatible representation that assigns amplitudes to the cells of the joint
frequency tables.

\subsection{Violations of commutative and distributive properties}

Quantum probabilities for sequential conjunctions violate the commutative
property. For example, referring to Figure \ref{projection}, consider the
quantum probability for conjunctive questions about events $X$ (e.g.,
democrat) and $U$ (e.g., moderate) again. The probability of agreeing to both
when question $X$ is queried first and question $U$ is asked second equals
$q(X"\sqcap U")=\left\vert M_{U}M_{X}\cdot\mathbf{S}\right\vert ^{2}=.2424,$
and the probability of yes to both in the opposite order is $q(U"\sqcap
X")=\left\vert M_{X}M_{U}\cdot\mathbf{S}\right\vert ^{2}=0.$ This dramatic
change in order happens in this case for the following reason. The initial
state $\mathbf{S}$ for the individual shown in the figure is orthogonal to the
vector $\mathbf{U}$. If this individual is initially asked about question $U$
(e.g., are you a moderate?), then there is zero probability of answering yes
to this first question (e.g., a person who likes to take a strong stand on
issues), and so the conjunctive probability is also zero. However, if the
individual is initially asked about the question $X$, then the initial state
$\mathbf{S}$ is negatively correlated to the vector $\mathbf{X}$ (e.g.
democrat), and its squared magnitude makes a reasonable probability of saying
yes and transiting from the $\mathbf{S}$ to the $\mathbf{X}$ state;
furthermore the $\mathbf{X}$ state (e.g. democrat) is positively correlated to
the $\mathbf{U}$ state (e.g., moderate), which then makes it possible to
transfer from $\mathbf{X}$ to $\mathbf{U}$ and answer yes to the second
question as well. In fact, it is well known that survey responses can be
manipulated by order \cite{McFarland_1982}, and similar `chaining' effects are
found in categorization \cite{Heit_1992}. Quantum probabilities for
disjunctions also violate the commutative property. For example, consider once
again Figure \ref{projection}. The quantum probability for the disjunction
question $(X\vee U) $ assuming that question $X$ is processed first equals
$q(X"\sqcup U")=$ $1-\left\vert M_{U^{\perp}}M_{X^{\perp}}\cdot S\right\vert
^{2}=.7273,$ and for the other order it is $q(U"\sqcup X")=$ $1-\left\vert
M_{X^{\perp}}M_{U^{\perp}}\cdot S\right\vert ^{2}=.4848.$ These differ because
$q(U")=0 $ for the latter order.

There is considerable direct evidence for order effects on the conjunctive
fallacy. \ In the first experiment of Gavansky and Roskos-Ewoldsen (1991),
participants rated the individual constituents before rating the conjunction
(producing the circles in Figure \ref{Gavansky_Fig}), and in the second
experiment the conjunction was rated first (producing the dots in Figure
\ref{Gavansky_Fig}). As can be seen, rating the conjunction first produced a
larger magnitude conjunction error. These results were replicated using random
assignment to two groups within a single study by Stolarz-Fantino et al.
(2003, Exp 2). When the conjunction came first, the mean probability rating
for the conjunction equaled .26 as compared to a mean of .18 for the low
event, and 57\% of the participants produced the error; but for the opposite
order the mean rating for the conjunction was .16 as compared to a mean of .14
for the low likelihood event, and only 31\% of the participants produced the error.\ 

The law of total probability is fundamental to Bayesian theory, but according
to quantum theory, it fails when incompatible events ever are involved. To see
how and why this happens, we return to Figure \ref{projection}. Consider the
probability for a question about event $W$ (e.g., whether or not a person is a
conservative). According to classic probability theory, a positive response to
this question can happen two mutually exclusive and exhaustive ways: the
person is an independent and a conservative $(Z^{\prime}\cap W^{\prime})$, or
the person is not an independent and a conservative $(\overline{Z}\cap
W^{\prime})$. So the total probability that a person is a conservative equals
$\mathbf{p}(W^{\prime})=\mathbf{p}((Z^{\prime}\cup\overline{Z})\cap
W)=\mathbf{p}(Z^{\prime})\cdot\mathbf{p}(W^{\prime}|Z^{\prime})+\mathbf{p}%
(\overline{Z})\cdot\mathbf{p}(W^{\prime}|\overline{Z}).$ Now let us reconsider
the quantum probabilities that we computed earlier for these events using
Figure \ref{projection}. When we first asked a question about $Z$ and then
asked about $W$, recall that we found $q(Z")=.0303$ and $q(W"|Z^{\perp
})=q(W|Z")=.50,$ and so the total probability is $q((Z"\sqcap W")\sqcup
(Z^{\perp}\sqcap W"))=$ $q(Z")\cdot q(W"|Z")+q(Z^{\perp})q(W"|Z^{\perp})=.50$.
But if we directly ask a person a question about event $W$, then we found
earlier that $q(W")=q((Z"\sqcup Z^{\perp})\sqcap W")=.6714$, which violates
the law of total probability! The reason that this happened is because the
initial state $\mathbf{S}$\ is very similar to the ray $W"$, but the initial
state $\mathbf{S}$\ is very dissimilar to the ray $Z"$ which must be reached
first by one of the two indirect routes from $\mathbf{S}$ passing through $Z"$
or $Z^{\perp\text{ }}$to $W".$ Violations of the law of total probability have
in fact been reported in some of earlier research \cite{BuseWangMog_2009}.
This violation of the law of total probability by quantum theory will turn out
to be one of the key ideas to explain the fallacies reviewed earlier. This
only happens when events are incompatible.

What determines the order for incompatible questions? This is an important
empirical issue. A working hypothesis is that when the individual events
differ greatly in terms of their likelihoods (e.g. for the Linda story, the
event feminist is very likely whereas the event bank teller is very unlikely),
then people start with the higher probability event. For the conjunction
question $(H\wedge L)$ this implies using the $(H"\sqcap L")$ conjunctive
sequence. For example, when asked the conjunction question regarding the Linda
story, we assume that the feminist event is processed before the bank teller
event. But for the disjunction question $(H\vee L)$, the relevant conjunction
question that needs to be considered is $(\symbol{126}H\wedge\symbol{126}L)$,
and $\symbol{126}L$ is more likely than $\symbol{126}H$. So the `start with
the higher probability' principle implies using the conjunctive sequence
$(L^{\perp}\sqcap H^{\perp}),$ which implies using the disjunctive sequence
$(L"\sqcup H")$. For example, when asked the disjunction question regarding
the Linda story, we assume that the not-bank teller event is processed before
the not-feminist event. Another factor that determines order of processing is
a cause - effect relation, i.e., if $C$ is the cause and $E$ is the effect,
then we assume $(C"\sqcap E")$. For example, when given the `increase tax and
reduce smoking problem', we assume that the 'tax' cause is processed first.

\section{Quantum Explanation of Judgment `Errors'}

The quantum model is essentially a similarity based approach to probability,
where similarity is determined by inner products of vectors in a
multidimensional space. Thus it is quite consistent with the finding that
typicality rating conjunction effects are highly correlated with conjunction
errors (Fact 14). \ In fact, it has already proved to be highly successful for
modeling typicality ratings for conjunctive and disjunctive concepts
\cite{Aerts&Gabora_2005}. But how do conjunction and disjunction errors arise
in the first place? We now turn to these more challenging questions.

\subsection{Conjunction error and its moderators}

Let us first consider a single conjunction fallacy (Fact 1). The state vector
$\mathbf{\psi}$ represents the memory state of the individual after reading
the story (which is based on both prior knowledge together with details about
the story). The projector, $M_{H}$ serves as a retrieval cue for retrieving
features related to the question about event $H$ (feminist); and similarly,
the projector $M_{L}$ serves as the retrieval cue for questions about event
$L$ (bank teller). Thus $M_{H}$ projects the Linda state $\mathbf{\psi}$\ onto
the high likelihood image of feminist, and $M_{L}$ projects the Linda state
$\mathbf{\psi}$ onto the low likelihood image of bank teller. According to the
`start with the higher probability' rule, the probability for the sequential
conjunction is $q(H"\sqcap L")=\left\vert M_{L}\cdot M_{H}\cdot\mathbf{\psi
}\right\vert ^{2}$, and the probability for the single event is
$q(L")=\left\vert M_{L}\mathbf{\psi}\right\vert ^{2}$. So how can we (the
theorists) tell whether or not the fallacy occurs? To do this, we (the
theorists, not the judge) need to express the single event probability in
terms of the conjunction probabilities using the quantum rules (see Appendix
for details):%
\begin{align}
q(L")  & =\left\vert M_{L}\mathbf{\psi}\right\vert ^{2}=\left\vert M_{L}\cdot
I\cdot\mathbf{\psi}\right\vert ^{2}\label{QConjEff}\\
& =\left\vert M_{L}\cdot(M_{H}+M_{H^{\perp}})\cdot\mathbf{\psi}\right\vert
^{2}\nonumber\\
& =q(H"\sqcap L")+q(H^{\perp}\sqcap L")+Int_{L}\nonumber
\end{align}%
\begin{equation}
Int_{L}=2\cdot\operatorname{Re}[(M_{L}M_{H^{\perp}}\mathbf{\psi})^{\dagger
}(M_{L}M_{H}\mathbf{\psi})].\label{Intereference}%
\end{equation}
Notice that the quantum probability (Equation \ref{QConjEff}) almost looks
like the law of total probability (Equation \ref{Law of Total Prob}), except
for the interference term, $Int_{L}$ (associated with event $L")$, which can
be positive, negative, or zero. This interference is the same mathematical
concept that is used to explain the classic two hole experiment with photons
in physics \cite{FeynLect}. If the interference term is zero, then quantum
probabilities satisfy the law of total probability and no conjunction error
occurs. Thus the model allows some people to be consistent with classic
probability theory. In particular, if $M_{H}$ and $M_{L}$ are compatible, then
this interference term is exactly zero (see Appendix). Thus interference only
occurs with incompatible events, and this explains why conjunction errors are
robust for questions about unrelated events $H_{1}$ and $L_{2}$ concerning
different people (Fact 13b). For this is exactly a situation in which it is
unlikely that a person has sufficient experience to form a compatible
representation, and must represent the situation with incompatible events that interfere.

To produce the conjunction `error' we require $Int_{L}<-q(H^{\perp}\sqcap
L")<0,$ and because $q(H^{\perp}\sqcap L")\geq0,$ this implies that the
interference must be sufficiently negative to produce a conjunction error.
This last result explains the fact the conjunction errors occur more
frequently with questions about mixed $H$ and $L$ events (Fact 12):
$q(H^{\perp}\sqcap L")$ must be small to produce the conjunction fallacy. Note
that if $H$ is a question about a high likelihood event, then $H^{\perp}$ is a
low likelihood quantum event, and $L"$ is also a low likelihood quantum event,
which makes $q(H^{\perp}\sqcap L")$ small, and so only a small negative amount
of interference is needed. This does not happen for the low -low case (because
$q(L_{1}^{\perp}\sqcap L_{2}")$ has one high component), or the high -high
case (because $q(H_{1}^{\perp}\sqcap H_{2}")$ has one high component), and so
the interference may be insufficient to produce the conjunction error in these
cases. In fact, the size of the conjunction error is bounded by the difference
between $q(H")\geq q(H"\sqcap L")\geq$ $q(L")$, and it shrinks to zero if
$q(H")=q(L")$ (see Appendix). This in fact matches the results shown in Figure
\ref{Gavansky_Fig}. There it can be seen that the conjunction error is present
only for mixed $H$ and $L$ events on the left wall, and it is absent for
events on the diagonal of the X-Y plane, where $q(A)$ is almost equal to
$q(B)$. Furthermore, consistent with Fact 12, only single conjunction errors
are predicted to occur in the high-low case, because the interference effect
is only produced for the $L"$ quantum event when sequentially processed in the
$(H"\sqcap L")$ order (see a later section for the double conjunction error issue).

How do we psychologically interpret this interference effect? Consider, for
example, Figure \ref{projection} once again. Suppose we compare the
conjunction probabilities $q(X"\sqcap U")$ and $q(X^{\perp}\sqcap U")$ with
the probability of the single event $q(U")$ given state $\mathbf{S}$ in the
figure. These calculations produce the following answers: $q(X"\sqcap U")$
$=\left\vert M_{U}M_{X}\cdot\mathbf{S}\right\vert ^{2}$ $=.2424$ and
$q(X^{\perp}\sqcap U")$ $=\left\vert M_{U}(M_{Y}+M_{Z})\cdot\mathbf{S}%
\right\vert ^{2}$ $=.2424$ but $q(U")$ $=\left\vert M_{U}\cdot\mathbf{S}%
\right\vert ^{2}=0$ and so $Int_{U}=-.4848$. The first term, $q(X"\sqcap U"),
$ is positive because $\mathbf{S}$ is negatively correlated with $\mathbf{X}$,
and $\mathbf{X}\,$\ is positively correlated with $\mathbf{U}$ in the figure,
and so the squared magnitude is positive. The second term is positive for the
same kind of reasoning. \ But $\mathbf{S}$ is orthogonal to $\mathbf{U}$ in
the figure. The psychological intuition behind this math is the following --
while it is possible to reach the conclusion $\mathbf{U}$ by way of thinking
first about $\mathbf{X}$ from state $\mathbf{S,}$ it is impossible to reach
this conclusion directly from state $\mathbf{S}$. In other words, the indirect
line of thought $\mathbf{S}\rightarrow\mathbf{X}\rightarrow\mathbf{U}$ has a
reasonable possibility even though there is no chance from the direct route
$\mathbf{S}\rightarrow\mathbf{U}$. You cannot see the conclusion $\mathbf{U}$
directly from state $\mathbf{S}$; but the indirect route (produced by asking
about question $X$ first) puts you in a state that makes you think of
something different, which then opens the possibility of reaching a conclusion
favoring yes to question $U$. For the Linda story, the judge cannot directly
imagine Linda as a bank teller; but if the judge first thinks about her as a
feminist, and then imagines her as a bank teller from this new feminist point
of view, it now seems more possible that she could be a bank teller. This
quantum explanation relates to both the availability and representativeness
heuristics. The representativeness heuristic comes into play when matching the
story to each question in terms of similarity, and the availability heuristic
comes into play when one question acts as a retrieval cue redirecting thinking
toward a different point of view.

The interference term can also be expressed as $Int_{L}=q(L")-[q(H"\sqcap
L")+q(H^{\perp}\sqcap L")].$ The first term, $q(L"),$ is the probability of
reaching a conclusion from a direct route (initial state to conclusion). The
bracketed term is the probability of reaching the same conclusion summed
across all indirect routes (through an incompatible set of eigenstates) to
that conclusion. Thus $Int_{L}$ is a quantity that we (the theorist) derive to
express the difference in probabilities caused by traveling the direct route
versus traveling a set of indirect routes, and different interference terms
can be derived depending on which set of indirect routes we compare to the
direct route. When the interference term is negative, that means that the
indirect routes have a greater chance of reaching the conclusion; and when the
interference term is positive, that means that the direct route has a greater
chance of reaching the conclusion. The interference term can be directly
estimated from experiments that request all three judgments $J(A)$, $J(A\wedge
B)$, $J(A\wedge\symbol{126}B)$.\footnote{This method requires strong
measurement assumptions for the judgment response.} This procedure was carried
out in the study by Wedell and Moro (2008), and using the data reported in
Table 2 from that article, we obtain the following interference estimates:
$-.36$ for the dice problem, and $-.55$ for the urn problem. The calculation
of the interference effect, $-.4848$, based on Figure \ref{projection} is an
example of a conjunction `fallacy' produced simply by using the inner products
(similarities) between vectors in the figure, and more exact results can be
obtained by adjusting these inner products. The inner products between vectors
are the key parameters for making exact predictions, and the model could be
fit to judgments using some type of multidimensional scaling algorithm. (This
would also require a more sophisticated response model.)

Note that $\operatorname{Re}[(M_{L}M_{H^{\perp}}\mathbf{\psi})^{\dagger}%
\cdot(M_{L}M_{H}\mathbf{\psi})]$ is the real part of the inner product between
two vectors, $M_{L}M_{H^{\perp}}\mathbf{\psi}$ and $M_{L}M_{H}\mathbf{\psi}$.
The first vector, $M_{L}M_{H^{\perp}}\mathbf{\psi}$ is the projection of the
state $\mathbf{\psi}$ (produced by the story) first on the $H^{\perp}$
subspace and then on to the $L"$ subspace; the second vector is the projection
of the same state $\mathbf{\psi}$ (created by the story) now on the $H"$
subspace and then again on the $L"$ subspace. For the Linda story,
$M_{B}M_{F^{\perp}}\mathbf{\psi}$ captures the features that match the Linda
story with a type of person who is first considered \textit{not} to be a
feminist and then considered also to be a bank teller; $M_{B}M_{F}%
\mathbf{\psi}$ captures the features that match the Linda story with a type of
person who is first considered to \textit{be} a feminist and then again
considered also to be a bank teller. Recall that an inner product is like a
correlation: If these two vectors match or are similar, then the inner product
will be positive; but if these two vectors mismatch or are dissimilar, then
the inner product will be negative; and if the two vectors are unrelated or
orthogonal, then the inner product will be zero. Although not many features
match between the Linda story and a person who is not feminist and a bank
teller; those that do match are likely to have some negative relation to those
that match a person who is a feminist and a bank teller, resulting in a
negative inner product and producing negative interference. More generally,
the relation between the features of the quantum events $H"$ and $L"$, as well
as their match to the story, are important for determining the size and
direction of interference. This is important for explaining Facts 10,11. The
interference depends on the inner product of projections on event subspaces,
and this inner product provides a principled way to understand the effects of
semantics and interdependence of events on conjunction errors. This inner
product also allows for effects of relationship between events that are
sometimes found (but not necessary) for conjunction errors (Fact 13a).

A similar analysis applies to the studies of the conjunction fallacy that
employ cause -- effect type of events. For example, suppose the two quantum
events are $C"$ (e.g., `increase cigarette tax') and $E"$ (e.g. `reduce
smoking'). \ The `cause first' order principle specifies that the prediction
for the conjunction is $q(C"\sqcap E"),$ and the prediction for the single
event $E"$ is%
\begin{align}
q(E")  & =\left\vert M_{E}\mathbf{\psi}\right\vert ^{2}=\left\vert M_{E}\cdot
I\cdot\mathbf{\psi}\right\vert ^{2}=\left\vert M_{E}\cdot(M_{C}+M_{C^{\perp}%
})\cdot\mathbf{\psi}\right\vert ^{2}\\
& =q(C"\sqcap E")+q(C^{\perp}\sqcap E")+Int_{E}.\nonumber
\end{align}
With negative interference produced by the sequential conjunctive judgment,
$Int_{E}<-$ $q(C^{\perp}\sqcap E")$, the quantum probabilities again produce a
conjunction fallacy $q(C"\sqcap E")>q(E")$. Again the psychological intuition
is the following. From the initial state, it is hard to imagine why teenage
smoking should decrease; but it is not hard to imagine a tax increase on
cigarettes, and once you imagine that, it is not hard to imagine a drop in
teenage smoking. If there is a strong causal relation, then $q(C"\sqcap E")$
$=q(C")\cdot q(E"|C")$ is large (because $q(E"|C")$ is large) and $q(C^{\perp
}\sqcap E")=q(C^{\perp})\cdot q(E"|C^{\perp})$ is small (because
$q(E"|C^{\perp})$ is small), and the conjunction fallacy is more likely to
occur. A positive conditional dependency between the cause and effect
increases the joint probability $q(C"\sqcap E")$ and decreases the joint
probability $q(C^{\perp}\sqcap E")$, which agrees with Fact 11. The
interference in this case equals $Int_{E}=\operatorname{Re}[(M_{E}M_{C^{\perp
}}\mathbf{\psi})^{\dagger}\cdot(M_{E}M_{C}\mathbf{\psi})]$. This means that
the inner product must be negative between (a) the projection first on the
cause absent followed by the effect, and (b) the projection first on the cause
present followed by the effect. In other words, the features produced by
situations associated with the cause absent and effect present are negatively
correlated with the features associated with the cause present and the effect present.

It is time to address the issue of double conjunction errors. Double
conjunction errors occur more \textit{frequently} for conjunctions that
contain two highly likely constituents. However, as can be seen in Figure
\ref{Gavansky_Fig}, double conjunction errors are not found using
\textit{means} even for $H\wedge H$ events. Quantum theory can only produce
zero or single conjunction errors. If $q(A")>q(B")$, then a single conjunction
error, $q(A"\sqcap B")>q(B")$, is possible (see Appendix). Double conjunction
errors obtained from a single rank ordering of a list of events can be
interpreted in one of two ways. First, they may simply be the result of
judgment error. This is a likely explanation for two reasons. One is that they
do not occur with the means after averaging out the error. Second, chance
errors from a single rank ordering of a list of events are very likely when
the event probabilities are nearly equal. In particular, if one assumes that
people correctly use the multiplicative rules of classic probability theory,
but base these calculations on noisy probability estimates, then more frequent
conjunction errors are predicted to occur by chance for the $H\wedge H$ case
\cite{Costello_2009}. A second, and possibly more interesting reason, is that
double conjunction errors may reflect an unusual situation in which the
formation of an entirely new unitized or configural concept emerges. More
formally, a new subspace $AB"$ is formed that corresponds to a projector which
cannot be decomposed into a product of the two projectors for the subspaces,
$A",B"$. The quantum concept of \textit{entanglement} has been used to
describe this new type of configuration \cite{Aerts&Gabora_2005}.

\subsection{Disjunction errors and unpacking effects}

Next let us consider the disjunction fallacy (Fact 2). Once again
$\mathbf{\psi}$ is the memory state following the Linda story, $M_{H}$ is the
retrieval cue or projector for feminist, and $M_{L}$ is the retrieval cue or
projector for bank teller. The quantum probability of the single event is
$q(H")=1-q(H^{\perp})$ and the quantum probability for the disjunction is
$q(L"\sqcup H")=1-q(L^{\perp}\sqcap H^{\perp}).$ Therefore, the disjunction
fallacy requires $q(H")>q(L"\sqcup H")$ $\ \mapsto q(H^{\perp})<q(L^{\perp
}\sqcap H^{\perp}).$ We (the theorists) can compare these predictions by
expanding $q(H^{\perp})$ like we did for $q(L")$ in Equation \ref{QConjEff}:%
\begin{align}
q(H^{\perp})  & =\left\vert M_{H^{\perp}}\cdot\mathbf{\psi}\right\vert
^{2}=\left\vert M_{H^{\perp}}\cdot(M_{L}+M_{L^{\perp}})\cdot\mathbf{\psi
}\right\vert ^{2}\label{DisEff}\\
& =q(L"\sqcap H^{\perp})+q(L^{\perp}\sqcap H^{\perp})+Int_{H^{\perp}%
},\nonumber\\
Int_{H^{\perp}}  & =\operatorname{Re}[(M_{H^{\perp}}M_{L}\mathbf{\psi
})^{\dagger}\cdot(M_{H^{\perp}}M_{L^{\perp}}\mathbf{\psi})].\nonumber
\end{align}
Using this result, we find that we require negative interference again,
$Int_{H^{\perp}}<-q(L"\sqcap H^{\perp}),$ to produce the disjunction effect.
As before we expect a single conjunction error when one event is high (in this
case it is $L^{\perp}$), and one event is low (in this case it is $H^{\perp}%
$). The psychological intuition in this case is the following. The disjunction
effect occurs when $q(L^{\perp}\sqcap H^{\perp}) $ becomes exaggerated, and
this happens because it is easy to think of Linda \textit{not} being a bank
teller (which leads one to say no), and once you start thinking about bank
tellers, it becomes harder to think about Linda as a feminist (which again
leads one to say no). But saying no to both of these questions leads to the
conclusion that the disjunction is false.

For example, consider once again Figure \ref{projection}. Suppose we compare
$q(X")$ with $q(U"\sqcup X").$ From our earlier calculations, we found that
$q(X)=\left\vert M_{X}S\right\vert ^{2}=.4848$. For the sequential disjunction
we obtain $q(U"\sqcup X")=1-q(U^{\perp}\sqcap X^{\perp})$ $=1-\left\vert
M_{Y+Z}\cdot M_{V+W}\cdot S\right\vert ^{2}=.4848.$ Thus we find
$q(X")=q(U"\sqcup X")$, which is still a disjunction error because using the
relations implied by Figure \ref{projection}, the probability of a yes to
question about $(U\wedge\symbol{126}X)$ is strictly positive. So according to
classic probability, if $p(U^{\prime}\cap\overline{X})>0$ then $p(U^{\prime
}\cup X^{\prime})$ $=p(U^{\prime}\cap X^{\prime})+p(\overline{U}\cap
X^{\prime})+p(U^{\prime}\cap\overline{X})$ $>p(U^{\prime}\cap X^{\prime
})+p(\overline{U}\cap X^{\prime})=p(X^{\prime}).$ Thus classic probability
requires $p(U^{\prime}\cup X^{\prime})>p(X^{\prime})$ with strictly greater
inequality in this example.

The \textit{real} challenge is to explain Fact 3 in which both conjunction and
disjunction fallacies occur within the same person and set of questions. This
requires $Int_{L}<-q(H^{\perp}\sqcap L")$ and $Int_{H^{\perp}}<-q(L"\sqcap
H^{\perp}),$ and these constraints need to be checked for feasibility. In the
appendix, we show that this set of constraints requires $q(L"\sqcap
H")=\left\vert M_{H}M_{L}\mathbf{\psi}\right\vert ^{2}$ $<\left\vert
M_{L}M_{H}\mathbf{\psi}\right\vert ^{2}=q(H"\sqcap L")$ which is consistent
with the theory when the events are incompatible. Psychologically speaking,
processing the high event first must facilitate retrieving a positive
conclusion to the conjunction more than processing the low event first. As we
have seen above, the sequential conjunction depends on the order.

The containment fallacy (Fact 4) can be explained using either Equation
\ref{QConjEff} or \ref{DisEff}, but it is more natural to use the former
because each question is actually about a single event. When shown a ski photo
and asked to the judge the likelihood that it came from Switzerland (question
$S$), the person answers yes to this event directly with quantum probability
$q(S)$. Similarly, when shown the ski photo and asked about the likelihood
that it came from Europe (question $E$), the person answers yes with quantum
probability $q(E)$. To compare these two probabilities, we (the theorists)
need to express $q(E)$ in terms of $q(S)$ as follows:
\begin{align*}
q(E)  & =|M_{E}\psi|^{2}=|M_{E}\cdot I\cdot\psi|^{2}\\
& =|M_{E}(M_{S}+M_{S^{\perp}})\cdot\psi|^{2}\\
& =q(S"\sqcap E")+q(S^{\perp}\sqcap E")+Int_{E}\\
& =q(S")\cdot q(E"|S")+q(S^{\perp}\sqcap E")+Int_{E}\\
& =q(S")\cdot1+q(S^{\perp}\sqcap E")+Int_{E}\\
& =q(S")+q(S^{\perp}\sqcap E")+Int_{E}.
\end{align*}
Once again we require negative interference $Int_{E}<-q(S^{\perp}\sqcap E")$
to produce the containment effect. The direct path from the state (produced by
the ski picture) to a positive conclusion about question $E$ (from Europe) is
low, but the indirect path from $S$ (from Switzerland) and then to $E$ (from
Europe) is very high, and so the interference is negative. This also requires
us to assume that people are using incompatible representations of these two
events, even though one question is about a subgroup of a larger group
referred to in the other question. This maybe a way of formalizing the gist
concept used in fuzzy trace theory to explain `class inclusion' illusions
\cite{Reyna_Brainerd_1995}.

Now consider unpacking effects (Fact 5). These effects can also be described
by interference between incompatible events \cite{BordKad1999}. The initial
finding by Rottenstriech and Tversky (1997) was that unpacking an event $D$
(death from murder) into a question about a likely event $S$ (killed by a
stranger) and another event ($\symbol{126}S$ killed by an acquaintance)
increases the judged probability when compared to the packed event. \ This
finding was explained by availability and formally incorporated as an
assumption into support theory, but quantum theory derives the effect using
the same line of reasoning as used for the conjunction error. First consider
the judgment for the packed quantum event $D"$ which we (the theorist not the
judge) expand in the same way as we did in Equation \ref{QConjEff}:%
\begin{align*}
q(D")  & =\left\vert M_{D}\cdot(M_{S}+M_{S^{\perp}})\cdot\mathbf{\psi
}\right\vert ^{2}\\
& =q(S"\sqcap D")+q(S^{\perp}\sqcap D")+Int_{D}\\
& =q(S")\cdot q(D"|S")+q(S^{\perp})\cdot q(D"|S^{\perp})+Int_{D}\\
& =q(S")\cdot1+q(S^{\perp})\cdot1+Int_{D}\\
& =q(S")+q(S^{\perp})+Int_{D}.
\end{align*}
The judgment for the implicit unpacked event is described by $q(S")+q(S^{\perp
}).$ In this case, the direct path to the conclusion for the packed event has
a lower probability than the sum of the indirect paths from the unpacked
events, producing negative interference: $Int_{D}<$ $2\cdot\operatorname{Re}%
[(M_{D}M_{S^{\perp}}\mathbf{\psi})^{\dagger}(M_{D}M_{S}\mathbf{\psi})]$ $<0.$
The negative interference implies that the projection of the initial state
first onto an acquaintance and then onto death is negatively correlated with
the projection of the initial state first onto stranger and then onto death.
This quantum interference explanation provides an alternative to support
theory for mathematically representing the effects of availability. The later
finding by Sloman et al. (2004) found that unpacking an event $D$ (death from
disease) into a question about a low likelihood event ($N$ death from
pneumonia) and a residual ($N^{\perp}$ diabetes, cirrhosis, and other
diseases) \textit{reduces} the judged probability compared to the packed
event. The quantum model agrees with the intuition provided by Sloman et al.
(2004) that when using an unlikely unpacked event and a residual, the indirect
paths produced by unpacking make it difficult to reach the conclusion, and now
it is easier to reach the conclusion directly from the unpacked event.
Although the latter find is contrary to the formalism of support theory, this
is still consistent with the quantum formalism, but now it produces positive
interference: $Int_{D}$ $=2\cdot\operatorname{Re}[(M_{D}M_{N^{\perp}%
}\mathbf{\psi})^{\dagger}(M_{D}M_{N}\mathbf{\psi})]$ $>0$. \ The positive
interference implies that the projection of the initial state first onto
pneumonia and then on to death is positively correlated with the projection of
the initial state first on to the residual (diabetes, cirrhosis, etc.) and
then onto death. Although support theory fails, the quantum model provides a
mathematically consistent way to formalize this interference effect using
positive or negative inner products.

There are at least two ways to explain the partitioning effect (Fact 6) using
the quantum model. One is to use interference as we did with the implicit
unpacking effect. However, a more convincing way is to use a quantum analogue
of Fox and Rottenstreich's (2003) `ignorance prior' (which can also be applied
to the implicit unpacking effect.) \ The original idea was based on the use of
a classical probability function $\mathbf{p}$ that assigns equal prior
probabilities to each alternative under consideration. Thus a focal event
receives greater probability in the case based representation (with only one
other comprehensive alternative) as compared to the class based partition
(with the comprehensive event broken down into several alternatives). The
quantum analogue uses a state vector $\mathbf{\psi}$ that assigns initial
amplitudes of equal magnitude to each alternative under consideration. This
results in the same `ignorance prior' effect by assigning a larger quantum
probability to the focal event in the case based partition as compared to the
class based partition.

\subsection{Averaging error, conditional fallacy, and inverse fallacy}

The averaging phenomena (Fact 9) easily can be explained by the quantum model.
This finding implies the that following inequalities are satisfied:%
\begin{align*}
q(L)  & =q(M\sqcap L)+q(M^{\perp}\sqcap L)+Int_{L}<q(M\sqcap L)\\
q(H)  & >q(H\sqcap M).
\end{align*}
This pair of inequalities follows directly from the earlier analyses. The
first inequality is satisfied as long as the interference, $Int_{L},$ is
sufficiently negative to produce a conjunction fallacy, and the second
inequality is always true for the quantum model when the high likelihood event
is processed first.

Now let us turn to Fact 7, the conditional fallacy. According to quantum
theory, the implication $J(H\mapsto L)$ is represented by L\"{u}der's rule.
But L\"{u}der's rule (like Bayes rule) cannot produce a conditional fallacy
because $q(H"\sqcap L")=q(H")\cdot q(L"|H")\geq q(L|H").$ However, there is a
simple alternative quantum explanation for this fallacy. Recall that
L\"{u}der's rule assumes that the judge first makes a transition from the
initial state (based on the story) to a state consistent with the antecedent
of the implication, $\mathbf{\psi\rightarrow\psi}_{H}$, before determining the
probability of the consequent of the implication. This projection only takes
place if the judge attends to the antecedent and accepts it as true. Otherwise
the projection fails to take place and the probability is based on the
projection of initial state $\mathbf{\psi}$ onto the consequent of the
implication, which implies that the judgment for the implication is based
simply on $q(L")$. Finally, if there is negative interference, then we obtain
$q(L")<$ $q(H"\sqcap L")$. This explanation agrees well with the findings of
Miyamoto et al. (1998) who found $J($the temperature remains below
$38^{o}F)\approx J($if it rains then the temperature remains below
$38^{o}F)\,<J($ it rains and the temperature remains below $38^{o}F).$ At the
same time, it can accommodate the findings of Tversky and Kahneman (1983) by
assuming that for this medical story, the truth of the antecedent `age is over
50' was attended and accepted as true, and a projection did occur before
determining the probability of the `heart attack' event.

A quantum explanation for the inverse `fallacy' is based on the idea that some
questions may be represented by simple rays (i.e., one dimensional vectors).
Consider two different questions, one is about an event represented by a ray
$A"$ (corresponding to the unit length vector $\mathbf{A}$); and another is a
question represented as another ray $B"$ (corresponding to the unit length
vector $\mathbf{B}$). Consider the quantum probability for the implication
$A\mapsto B$ computed from L\"{u}der's rule. First we project the initial
state $\mathbf{\psi}$ onto the ray $A"$ and normalize:
\[
\mathbf{\psi}_{A}=M_{A}\mathbf{\psi}/\left\vert M_{A}\mathbf{\psi}\right\vert
=(\mathbf{A}\cdot\mathbf{A}^{\dagger}\cdot\mathbf{\psi})/\left\vert
\mathbf{A}\cdot\mathbf{A}^{\dagger}\cdot\mathbf{\psi}\right\vert =\mathbf{A}.
\]
As can be seen in this special case of events based on rays, the state simply
changes from the vector $\mathbf{\psi}$ to the vector $\mathbf{A}$. Next we
compute the quantum conditional probability for one ray given another ray:%
\begin{align*}
q(B"|A")  & =\left\vert M_{B}\mathbf{\psi}_{A}\right\vert ^{2}=\left\vert
\mathbf{B}\cdot\mathbf{B}^{\dagger}\cdot\mathbf{A}\right\vert ^{2}\\
& =\left\vert \mathbf{B|}^{2}\cdot|\mathbf{B}^{\dagger}\cdot\mathbf{A}%
\right\vert ^{2}=\left\vert \mathbf{B}^{\dagger}\cdot\mathbf{A}\right\vert
^{2}.
\end{align*}
As can be seen from the above, this is just the squared magnitude of the inner
product between vectors $\mathbf{A}$ and $\mathbf{B}$. However, if we repeat
this procedure in the opposite direction for $B\mapsto A$ we obtain
$q(A"|B")=\left\vert \mathbf{A}^{\dagger}\cdot\mathbf{B}\right\vert ^{2}$.
Thus whenever events $A"$, $B"$are simply rays, we obtain the equality%
\[
q(A"|B")=\left\vert \mathbf{A}^{\dagger}\cdot\mathbf{B}\right\vert
^{2}=\left\vert \mathbf{B}^{\dagger}\cdot\mathbf{A}\right\vert ^{2}=q(B"|A").
\]
It is important to remember that this equality is \textit{not} true in general
for subspaces that have dimensions greater than one, because in this case
conditional probability does not reduce to a single inner product. Thus
quantum theory can explain the inverse fallacy whenever the questions are
represented as simple one dimensional rays. This can happen if a person relies
on an oversimplified vector space representation to answer questions. Suppose
an individual is asked a question such as `disease present (or absent) given
test result positive (or negative).' In this case, a person may represent the
problem using two incompatible sets of projectors operating within the same
two dimensional space: one set based on eigenvectors $\{\mathbf{D}%
,\mathbf{D}^{\perp}\}$ representing disease present or absent, and another
based eigenvectors $\{\mathbf{T},\mathbf{T}^{\perp}\}$ representing a positive
or negative test. Given this oversimplified representation, the person would
produce an inverse fallacy because $q(D"|T")=\left\vert \mathbf{D}^{\dagger
}\cdot\mathbf{T}\right\vert ^{2}$ $=\left\vert \mathbf{T}^{\dagger}%
\cdot\mathbf{D}\right\vert ^{2}=q(T"|D")$. Note, however, that this
oversimplified representation of events could not be used to answer questions
about other diseases and/or combinations of test results, which would require
a higher dimensional space.

\subsection{Order effects and response mode}

Finally we consider the order effect that occurs when conjunctions are rated
first as compared to last (compare the circles with the dots in Figure
\ref{Gavansky_Fig}). Suppose the constituent events, $(L,$ $H)$ are rated
separately first. Then either one of these two estimates can be used later to
estimate the conjunctive probability. If the person selects the $q(L")$
estimate, then the conjunction can be computed from $q(L")\cdot
q(H"|L")>q(L")$ and no conjunction error can occur. If the person selects the
$q(H")$ estimate, then the conjunction is computed from $q(H")\cdot
q(L"|H")=q(H"\sqcap L")$ which can exceed $q(L")$ to produce a conjunction
error. So if some proportion start with each estimate, then the conjunction
error is reduced by this proportion. This reduction does not happen in the
reverse order because in this case, the `start with the higher probability
event first' rule applies and the conjunction is always computed from
$q(H"\sqcap L")$ which produces conjunction errors. This order effect can also
explain why ratings produce fewer errors than rank orders, because the latter
does not request any estimates of the constituents ahead of time. This
explains Fact 15.

\section{Comparison with Previous Theories}

A brief comparison of the quantum model with three previous theories for
conjunction and disjunction errors (and related findings) is presented below.
Table 1 provides a summary indicating whether or not each theory can explain
each finding (y=yes, n=no, u=unknown).

\bigskip%

\begin{tabular}
[c]{|c|c|c|c|c|c|c|c|c|c|c|c|c|c|c|c|}\hline
\multicolumn{16}{|c|}{Table 1: Summary of each explanation for 15 major
findings.}\\\hline
& \multicolumn{15}{|c|}{Fact}\\\hline
& {\normalsize 1} & {\normalsize 2} & {\normalsize 3} & {\normalsize 4} &
{\normalsize 5} & {\normalsize 6} & {\normalsize 7} & {\normalsize 8} &
{\normalsize 9} & {\normalsize 10} & {\normalsize 11} & {\normalsize 12} &
{\normalsize 13} & {\normalsize 14} & {\normalsize 15}\\\hline
{\normalsize R} & {\normalsize y} & {\normalsize y} & {\normalsize y} &
{\normalsize y} & {\normalsize n} & {\normalsize n} & {\normalsize y} &
{\normalsize n} & {\normalsize n} & {\normalsize n} & {\normalsize y} &
{\normalsize y} & {\normalsize n} & {\normalsize y} & {\normalsize y}\\\hline
{\normalsize A} & {\normalsize y} & {\normalsize y} & {\normalsize y} &
{\normalsize n} & {\normalsize n} & {\normalsize n} & {\normalsize n} &
{\normalsize n} & {\normalsize y} & {\normalsize n} & {\normalsize n} &
{\normalsize y} & {\normalsize y} & {\normalsize n} & {\normalsize n}\\\hline
{\normalsize M} & {\normalsize y} & {\normalsize n} & {\normalsize n} &
{\normalsize n} & y & {\normalsize y} & {\normalsize u} & {\normalsize n} &
{\normalsize y} & {\normalsize n} & {\normalsize n} & {\normalsize y} &
{\normalsize y} & {\normalsize y} & {\normalsize n}\\\hline
{\normalsize Q} & {\normalsize y} & {\normalsize y} & {\normalsize y} &
{\normalsize y} & {\normalsize y} & {\normalsize y} & {\normalsize y} &
{\normalsize y} & {\normalsize y} & {\normalsize y} & {\normalsize y} &
{\normalsize y} & {\normalsize y} & {\normalsize y} & {\normalsize y}\\\hline
\end{tabular}

\begin{center}
Note: R = representativeness, A = averaging, M = memory, Q = quantum.
\end{center}

Tversky and Kahneman (1983) initially argued that many judgment errors result
from the use of the representativeness heuristic, which relies on the
similarity between the features generated by the story and the features
entailed by the event in question. This idea agrees well with the quantum
probability model. The representativeness heuristic was initially used to
explain Facts 1,2,3,4,7. It was never applied to Facts 5,6,8,9,10. It is at
least consistent with Facts 11,12. This explanation fell into disfavor mainly
because of Fact 13, in which conjunction errors occurred almost equally often
for related and unrelated events. For example, suppose $L_{1}$ =`Linda is a
bank teller' and $H_{2}$ = `Bill is an accountant.' If $J(L_{1}\wedge
H_{2})>J(L_{1})$, then it is argued that this result cannot arise from
representativeness -- there is no single stereotype or prototype associated
with the conjunction in this case. Finally, representativeness agrees well
with Fact 14. Fact 15 is consistent with the idea that heuristic thinking is
more likely to be evoked by ranking procedures, and analytic thinking is more
likely to be evoked by rating scales, but this is a bit post hoc. One
outstanding problem with the notion of representativeness is that it lacks a
clear or rigorous formalization, which makes it difficult to determine exactly
what it predicts or does not predict \cite{Gigerenzer_1996}. Support theory is
such a formalization, but it is based on availability rather than
representativeness \cite{Tversky-Koehl_Support_2004}. It was devised to
explain Facts 5 and 6, but it has not been systematically applied to the other
facts. However, in their discussion section, Tversky and Koehler (2004) point
out that one could use support theory to model the conjunction fallacy by
viewing a question about event $B$ as a `packed' version corresponding to the
unpacked event $(B\wedge F)$ $\vee(B\wedge\symbol{126}F)$. While the unpacked
event must produce a greater judged probability than a conjunction that it
contains, the packed version could produce a smaller judged probability than a
conjunction it contains. This is closely related to the explanation from
quantum theory which uses the expansion in Equation \ref{QConjEff} in a
similar manner.

A second major competing explanation is that people average the evidence
provided by each separate event \cite{Wyer_1976}, and this explanation is
gaining support \cite{Nilsson_2008}. According to this theory, a person
assigns a subjective probability to each component event that may appear in a
combination. For example, $S(A)$, $S(B)$, and $S(C)\,$\ denote subjective
probabilities assigned to single events $A,B,C,$ which may also appear in
questions about combinations of these events. Importantly, these subjective
probabilities are assigned independent of the other events for which they are
combined. For example, the same $S(A)$ is used for event $A,$ $(A\vee B), $
and $(A\wedge C)$. The evidence for an event composed of the two individual
events $(A,B)$ is formed by an average: $r(AandB)$ denotes the weighted
average when asked about the conjunction $(A\wedge B)$, and $r(AorB) $ denotes
the weighted average for the disjunction question $\left(  A\vee B\right)  $.
For any two events, one is more likely than the other, and so we will let
event $H$ represent the more likely event and $L$ represent the less likely
event of the pair. Then the averages for the conjunction and disjunction are%
\begin{align}
r(LandH)  & =S(L)+w_{1}\cdot(S_{H}-S_{L})=(1-w_{1})\cdot S(L)+w_{1}\cdot
S(H)\label{Avg model}\\
r(HorL)  & =S(H)+w_{2}\cdot(S_{L}-S_{H})=(1-w_{2})\cdot S(H)+w_{2}\cdot
S(L).\nonumber
\end{align}
Different weights ($0<w_{i}<1$) are used for different types of questions
about combinations of events. How does one determine these weights? For
conjunction tasks, it is assumed that people anchor on the lower probability,
and then adjust upward toward the higher probability producing the processing
order $L\&H$; for disjunction, people anchor on the higher probability, and
adjust downward toward the lower probability producing the processing order
$HorL$. This processing order presumably causes more weight to be placed on
the lower probability event for conjunctions; and it causes more weight to be
placed on the higher probability event for disjunctions. This idea of
anchoring and adjustment is analogous to the processing order assumptions used
in the quantum model. Alternatively, a geometric average has been used to
model these effects \cite{Abelson_conj_1987}, but this model makes the same
ordinal predictions as the averaging model because the log of the geometric
average equals the arithmetic average. So far this model has not been applied
to implications and conditional probabilities, and so it remains unclear how
to do this. The averaging model readily explains Facts 1,2,3. It has never
been applied to Facts 4,5,6,7,8 and it is unclear how it would explain these.
It agrees very well with Fact 9. But the averaging model has major problems
with Facts 10 and 11 because it ignores dependencies between events. A
deterministic interpretation of the averaging model also has problems with
Fact 12: strictly speaking, a single conjunction error should \textit{always}
occur for unequal events because the average always falls between the two
unequal values being averaged. The same problem is true for disjunction
errors. This is a problem because conjunctions and disjunction errors do not
\textit{always} occur. For some people they never occur and for some pairs of
questions they never occur. This problem can fixed partly by assuming there is
noise in the judgment process, in which case zero and double
conjunction/disjunction errors can occur by chance \cite{Nilsson_2008}. This
implies that `correct' judgments (i.e., zero errors) are caused by random
noise and single conjunction errors are the norm for all people and for all
pairs of questions. This model has no explanation for the last two facts 14,
15. One final criticism of the averaging model is its lack of coherence -- the
weights assigned to implications are not constrained by assignments to
conjunctions, and the model is unable to handle mutually exclusive events. For
example, suppose $J(A)>J(\symbol{126}A)$; then the averaging model implies
that $J(A)>J(A\vee\symbol{126}A)>J(A\wedge\symbol{126}A)>J(\symbol{126}A)$ and
so conjunction and disjunction errors should be frequent with mutually
exclusive events. Recent evidence indicates that conjunction and disjunction
errors are greatly reduced when they are formed from mutually exclusive events
\cite{WolfeReyna_2009}.

A third major explanation is that probability judgment errors result from a
feature based memory retrieval process \cite{DoughertyGettysOdgen_1999}. Two
types of explanations were proposed, one for judgments based on stories
(vignettes), and the other for judgments based on training examples, but all
of the studies in our review are based stories (vignettes), and so we limit
our discussion to the first explanation. According to this model, information
about the story is stored in a memory trace (column) vector denoted
$\mathbf{T}$. The coordinates of this memory vector represent positive or
negative feature values related to the story, and zeros are assigned to
features unrelated to the story. A single question $A$ is represented by a
probe (column) vector of the same length, $\mathbf{P}_{A}$, with values
assigned to features related to both the question and the story, and zeros
otherwise. Retrieval strength (echo intensity) to a question is determined by
the inner product between the memory trace vector and the question probe
vector, $I_{A}=[(\mathbf{P}_{A}^{\prime}\cdot\mathbf{T})/N_{A}]^{3}.$ Note
that the inner product is normalized by dividing it by a number, $N_{A}$, that
depends on the number of nonzero elements in the question probe vector, and
the inner product is cubed to quench small echoes and exaggerate strong
echoes. Frequency or relative frequency judgments are assumed to be
proportional to echo intensity (which requires the intensity to be
non-negative). The use of feature vectors to represent the memory state for
stories and the probe for questions, as well as the use of inner products to
determine probability judgments, is conceptually close to quantum probability
theory (except that the inner products would be squared rather than cubed to
produce quantum probabilities). Conditional probabilities are estimated by a
two-part process of first retrieving traces similar to the probe, and then
applying a threshold that retains only traces with sufficiently strong echos.
The threshold mechanism is not part of Bayes' rule used in classic theory nor
L\"{u}der's rule used in quantum theory, although it reduces to Bayes' rule as
a special case for very low thresholds. However, for conjunction questions,
the memory retrieval model does \textit{not} assume a sequence of two
retrievals (one retrieval for the first constituent and a second retrieval
conditioned on the first for the other constituent, as assumed by the quantum
model), and so it does not make use of its conditional probability mechanism
for these types of questions. Instead, a conjunctive question $H\wedge L$ is
represented by a single conjunctive probe, which is the sum
(concatenation)\footnote{Dougherty et al. described the conjunctive probe as
the concatenation of two minivectors, but this is the same as summing two
non-overlapping vectors. If $P_{H}$ is a row minivector for $H$ with length
$N_{H},$ and $P_{L}$ is a row minivector for $L$ with length $N_{L},$ and
$0_{N}$ is a row vector of $N $ zeros, then $[P_{H}|P_{L}]=[P_{H}|0_{N_{H}%
}]+[0_{N_{L}}|P_{L}].$} of the probes used for the two constituent question
vectors, $\mathbf{P}_{H\And L}=\mathbf{P}_{H}+\mathbf{P}_{L}$. The echo
intensity of this conjunction probe produces something akin to an average,
\begin{align}
\sqrt[3]{I_{H\&L}}  & =(\mathbf{P}_{H\&L}^{\dagger}\cdot\mathbf{T}%
)/N_{H\&L}=[(\mathbf{P}_{H}^{\dagger}+\mathbf{P}_{L}^{\dagger})\cdot
\mathbf{T}]/N_{H\&L}\label{mem model}\\
& =(\mathbf{P}_{H}^{\dagger}\cdot\mathbf{T})/N_{H\&L}+(\mathbf{P}_{L}%
^{\dagger}\cdot\mathbf{T})/N_{H\&L}\nonumber\\
& =\frac{N_{H}}{N_{H}+N_{L}}\cdot\sqrt[3]{I_{H}}+\frac{N_{L}}{N_{H}+N_{L}%
}\cdot\sqrt[3]{I_{L}}.\nonumber
\end{align}
In particular, the above implies $\sqrt[3]{I_{L}}<\sqrt[3]{I_{H\&L}}%
<\sqrt[3]{I_{H}}$ , which by monotonicity implies $I_{L}<I_{H\&L}<I_{H}$. In
short, the memory retrieval model explains the conjunction error (Fact 1) in
the same way as the averaging model. The memory model has an additional
advantage because it provides a similarity based mechanism for determining the
echo intensities for each question. This is useful for relating the features
in the memory model to the semantic aspects of story and a single event.
Dougherty et al. (1999) did not address disjunction errors (Facts 2,3,4), and
it is unclear how the model applies to these results; but more recent
extensions have been formulated to acount for `unpacking' effects
\cite{ThomasDougherty_2008}.The memory model includes a mechanism for
computing conditional probabilities, which could be (but has not been) used to
explain Fact 7. If this conditional mechanism reverses direction, it would
produce a inverse fallacy, but the theory does not explain why people
sometimes reverse the conditional, and so it does not really explain Fact 8.
Like the averaging model, the memory model can accommodate Fact 9, but at this
point it has no explicit mechanism for explaining event dependencies and
conjunction effects (Facts 10, 11). The latter problem arises from the fact
that the features are defined for each event separately, and then they are
simply added (concatenated) together for conjunctions. This is the same
problem that arises with the averaging model. Also like the averaging model,
the memory retrieval model always predicts single conjunction errors, and like
the averaging model, some type of error or sampling variability is required to
explain the occurrence of zero or double conjunction errors (Fact 12). The
memory retrieval model can accomodate related and unrelated conjunction errors
using the summation of individual vectors to represent conjunctions (Fact 13).
The memory model, being based on an inner product measure of similarity is
also consistent with Fact 14, but it does not address response mode or order
effects (Fact 15).

The summary shown in Table 1 indicates that the quantum model provides a more
comprehensive account of conjunction and disjunction errors and closely
related phenomena in comparison with the other three theories. We do
\textit{not} conclude that the quantum model is superior in general to any of
these other theories -- they have been applied to many other phenomena beyond
conjunction and disjunction errors that are not covered here (such as
inference problems and base rate neglect). We simply wish to conclude that the
quantum model provides a viable and promising new approach to understanding
conjunction and disjunction errors and related phenomena. Future work will
extend the model to inference \cite{Buse&True_2009}.

\subsection{Model complexity and testability}

Quantum probability contains classic probability as a special case, and
therefore it is more complex than classic probability theory.\footnote{Quantum
theory is not the only way to generalize classic probability theory. An
alternative is to describe events as open sets from a topology which replaces
set theoretic complementation with a less stringent pseudo -complementation
\cite{Narens_2009}. The latter theory has been used to explain `upacking'
effects.} But so are the other explanations for judgment fallacies, and this
criticism is not unique to quantum theory. It is hard to say at this point
which of the competing explanations is more complex. For example, we don't
know at this point if the quantum or memory retrieval model is more complex.
We can, however, point to places where this quantum model makes clear testable
predictions for future research.

The quantum model must predict that single conjunction errors only occur when
one event has a high likelihood and the other has a low likelihood, and zero
conjunction errors should occur when the two probabilities are equal (except
for response errors). These predictions agree well with the results presented
in Figure \ref{Gavansky_Fig}. The same prediction must hold for disjunction
errors, which is also empirically supported. Quantum probability theory cannot
produce a double conjunction error (except by response error). If future
research proves that this phenomena is systematic and replicable, then the
quantum model needs to be extended to provide a new principle for forming
conjunctive concepts that are unitized and can no longer be decomposed into
parts. Quantum probability theory predicts no conjunction or disjunction
errors for complementary events $A,\symbol{126}A$ (except those produced by
response errors), whereas the averaging model predicts they will be as robust
as ever. In fact, Wolfe and Reyna (2009) report a reduction but not
elimination of conjunction and disjunction errors for complementary events as
compared to pairs that overlap in probability. The quantum model must predict
that if events $A$ and $B$ are found to be compatible in a study on
conjunctions, then these same two events $A,B$ are predicted to produce no
disjunction errors either. The quantum model also predicts that if events
$A,B$ are found to produce a conjunction error, then they must be
incompatible, and therefore the judgments of the joint probabilities must
change depending on the order that they are processed. Although this has not
been directly tested yet, other evidence for order effects consistent with
this prediction was presented. Finally, the theory makes novel predictions for
conjunctions, disjunctions, and conditionals involving more than two events,
and these predictions also can be derived directly from the general principles
without adding any new assumptions. An important step for future work is to
include a more complete choice response model using quantum theory, and some
initials steps in this direction have been made \cite{BusQD}. This addition is
critical for deriving quantitative and probabilistic rather than qualitative
and deterministic predictions from the model.

\section{Fuzzy Reasoning Under Uncertainty}

Both classic (Kolmogorov) and quantum (von Neumann) probability theories are
based on a coherent set of principles. In fact, classic probability theory is
a special case of quantum probability theory in which all the events are
compatible, which generates a simple Boolean algebra of events. Incompatible
events produce a more complex `partial' Boolean algebra of events
\cite{Hughes_1989}. So why do we need to use incompatible events, and isn't
this irrational? In fact, the physical world obeys quantum principles and
incompatible events are an essential part of nature. Clearly there are many
circumstances where everyone agrees that the events should be treated
classically (such as random selection of balls from urns or dice throwing).
But incompatible events may be essential for understanding our commonly
occurring but nevertheless very complex human interactions. For instance, when
trying to judge something as uncertain as winning an argument with another
person, the likelihood of success may depend on using incompatible
representations that allow viewing the same facts from different perspectives.

The use of incompatible events introduces a new and potentially useful concept
to cognition, which is called a \textit{superposition} state. This concept is
fundamentally different than the concept of a \textit{mixed} state used in
classic Bayesian probability theory. Consider the events depicted in Figure
\ref{projection} again. Suppose that a voter knows that she definitely will
\textit{not} vote for the independent candidate ($Z$), and therefore she must
vote for either the democrat ($X$) or the republican ($Y$). If she is in a
classic mixed state at a moment before the decision, then she is exactly in
one state (favoring a vote for democrat) or the other (favoring a vote for
republican) and not both at that moment. If she cannot consciously say which
state exists at that moment, she could express the probability that
\textit{the} true state is one that favors democrat or republican, but a
precise state \textit{does} exist at the moment before the decision, and the
final act of voting simply records the immediately preexisting but possibly
unknown state. If she is in a quantum superposition state, then she is
\textit{not} exactly in a state favoring the democrat, \textit{not} exactly in
a state favoring the republican, and not \textit{exactly} in both states
immediately before the vote is cast. She cannot verbally say (with respect to
democrat and republican) exactly what the state is before the vote is case,
because no clear state exists with regard to these two outcomes. Perhaps it is
best characterized as a \textit{fuzzy} state with regard to democrat and
republican before the measurement. The act of voting creates a clear and
specific state (e.g. the person becomes a democratic voter after casting her
vote). The superposition state better matches the well accepted idea that
preferences are constructed on the spot for the purpose of making judgments or
taking actions rather that being determined a priori \cite{PayneBettJohn_1992}.

What are the behavioral implications of this distinction between mixed versus
superposition states? Suppose the following conditional probabilities are
known to be true for \textit{both} the classical and quantum systems. (These
probabilities match the situation depicted in Figure \ref{projection}.) If the
person is not an independent, then there is a .50 chance that she votes
democrat or republican; if the person votes democrat, then there is a .50
chance she claims to be a moderate; if the person votes republican, then there
is a .25 chance she that she claims to be a moderate. Now according to classic
probability theory, if the person tells us she is not an independent, then she
is going to vote for a democrat or republican (exclusively), and so the total
probability that she will claim to be moderate (if asked right before casting
her vote) must be $(.50\cdot.50)+(.50\cdot.25)=.375$. According to the quantum
system, if we learn that the person is not independent, then the person is in
the superposed state $\mathbf{A}$ in Figure \ref{projection} (which is neither
democrat nor republican), and the probability of being a moderate from this
state (before casting the vote) turns out to be exactly zero (the vector
$\mathbf{A}$ is orthogonal to the vector $\mathbf{U}$ for moderate in Figure
\ref{projection}) ! So we see that superposition states do not behave the same
way as classic mixed states.

The superposition concept is related to other theories of fuzzy reasoning.
Fuzzy set theory has been useful in psychology for representing vague verbal
expressions used in natural language \cite{WallstenBudescu_1986}. For
instance, a vague expression such as `Tom is short' is represented in fuzzy
set theory by a membership function that assigns membership values to the
levels of a meter scale. This corresponds to the quantum superposition state
with probability amplitudes assigned to eigenvectors associated with the
different levels on the meter scale. Quantum theory can enhance fuzzy set
theory by providing a more powerful formalism for evaluating complex
combinations of expressions. Fuzzy trace theory has been useful for
understanding how people use gist versus precise representations to reason
under uncertainty \cite{Reyna_Brainerd_1995}. The superposition principle
provides a natural way to represent a `gist' state as a superposition over
precise values. Consider the gist `this program could save lives.' This gist
could be represented as a classical mixed state by a probability distribution
over number of lives saved. Alternatively, it could be represented as a
superposition state by an amplitude distribution over eigenvectors
representing number of lives saved. The classic representation assumes that
exactly one and only one number saved is the correct hypothesis, but we don't
know which one it is, and we assign a probability to each hypothesis; whereas
the quantum representation rejects the assumption that exactly one number is
correct, and instead retains a fuzzy representation. As pointed out above,
these two different representations of uncertainty can produce very different
predictions for decision making behavior \cite{LaMura_2009}. Quantum theory
could be useful for formalizing some of the principles of fuzzy trace theory.

In summary, we argue that it is important to introduce a distinction between
compatible and incompatible representation of events when describing human
judgments. More accurately, we should say `re-introduce' this distinction,
because Bohr actually got the idea of complementarity from William James.
Human judges may be capable of using either compatible or incompatible
representations, and they are not constrained or forced to use just one. The
use of compatible representations produces judgments that agree with the
classic and Bayesian laws of probability, whereas the use of incompatible
representations produces violations. But the latter may be necessary to deal
with deeply uncertain situations (involving unknown joint probabilities),
where one needs to rely on simple incompatible representations to
\textit{construct} sequential conjunctive probabilities \textit{coherently}
from quantum principles. In fact, both types of representations, compatible
and incompatible, may be available to the judge, and the context of a problem
may trigger the use of one or the other \cite{Reyna_Brainerd_1995}. More
advanced versions of quantum probability theory (using a Fock space, which is
analogous to a hierarchical Bayesian type model) provide principles for
combining both types of representations \cite{Aerts_JMP_2009}.

\section{Concluding Comments}

It is worthwhile to briefly consider how a quantum framework for cognition
relates to the many alternative models which have been explored in the last
few decades. A quantum approach is most closely aligned to Bayesian
approaches. In the latter, inference is guided by the updating of
probabilities through Bayes's rule. Bayesian models have been successfully
applied to many aspects of cognition, such as similarity
\cite{Tennenbaum&Griffiths_2001}, reasoning \cite{OaksChater_2009}, language
processing \cite{GriffSteyvTen_2007}, and categorization \cite{Anderson_1991}.
As noted earlier, quantum probability theory has an analogue of Bayes's rule,
called L\"{u}der's rule, which is used to update inferences. Of course, there
are other key differences between quantum and Bayesian approaches, notably the
order-dependence of operations in the former which are order-independent in
the latter (such as conjunction and disjunction). In this sense, a quantum
approach can be thought of as a generalized Bayesian approach. There are also
relations between quantum approaches and other computational paradigms for
modeling cognition. For example, quantum computing models \cite{NielCh2000}
provide parallel processing capabilities as championed by connectionist models
\cite{Rum&McClell_1986}, but at the same time they are able to take advantage
of condition - action procedures that match classical production rule systems
\cite{AndersonBook_1993}. However, learning is a new challenge for quantum
information processing systems. On the basis of some of the relevant
computational examinations in the literature \cite{BoucherDienes_2003}, we can
provisionally suggest that the main difference between quantum models and such
alternatives would be the order dependence of operations in quantum
information processing. In the field of decision making, we have examined the
dynamics of quantum models in more detail and have found evidence for
interference effects which sharply contrast with that of corresponding Markov
models \cite{PothosBusemeyer_2009}. Such results indicate that a quantum
approach to modeling cognition will produce very distinct computational predictions.

In closing, quantum probability theory is brand new for psychologists,
cognitive scientists, and decision scientists. It may seem to be a strange
idea at first, but once familiar with it, the theory has some appealing
properties for cognition in particular and psychology in general. On the one
hand, quantum probability provides a powerful and coherent framework for
modeling human judgments that compares with classic (Kolmogorov) probability
theory. On the other hand, it provides a geometric (similarity) based approach
to probability that provides new psychological concepts for reasoning under
uncertainty, such as incompatible representations of events, superposition
states of beliefs, and interference among paths to conclusions. In this
article we demonstrate that quantum probability theory provides a viable and
promising explanation for conjunction and disjunction fallacies and closely
related phenomena. In future work we plan to apply the model to inference
tasks which in the past have been explained using Bayesian modeling frameworks
\cite{Buse&True_2009} .

\section{Author Notes}

This material is based upon work supported by the National Science Foundation
under Grant No. 0817965. We thank Rich Shiffrin for his thought provoking
questions about quantum versus classic probability theories.

\bibliographystyle{IEEEtran}

\section{\bigskip Appendix}

\subsection{\bigskip Derivation of quantum probabilities from projectors}

In quantum theory, an eigenvector can be a complex vector. We do not to limit
our theory to real vectors, and all derivations allow for complex vectors, but
the examples use only real vectors for simplicity. If $\mathbf{V}$ represents
an $n\times1$ column vector with a complex coordinate $v_{k}=x+i\cdot y$ in
row $k$, then $\mathbf{V}^{\dagger}$ is the Hermitian transpose, which changes
the column vector into a row vector replacing the original coordinate with its
conjugate $v_{k}^{\ast}=x-i\cdot y$ in row $k$ ($i=\sqrt{-1}$, $x,y\in
\operatorname{real}$). Real numbers are a special case with $y=0$. An
amplitude $\psi_{j}$ may also be complex, and so the squared magnitude is
defined as $\left\vert \psi_{j}\right\vert ^{2}=\psi_{j}\cdot\psi_{j}^{\ast}.$
The inner product between two complex vectors $\mathbf{V},\mathbf{W}$ is a
scalar (a complex number) whose value depends on the order: $\mathbf{V}%
^{\dagger}\cdot\mathbf{W}=\sum_{i}v_{i}^{\ast}\cdot w_{i}$ and $\mathbf{W}%
^{\dagger}\cdot\mathbf{V}=\sum_{i}w_{i}^{\ast}\cdot v_{i}$ $=(\mathbf{V}%
^{\dagger}\cdot\mathbf{W})^{\ast}$ but $|\mathbf{W}^{\dagger}\cdot
\mathbf{V}|=|\mathbf{V}^{\dagger}\cdot\mathbf{W}|$. Two vectors $\mathbf{V}%
,\mathbf{W}$ are orthogonal if $\mathbf{V}^{\dagger}\cdot\mathbf{W}=0,$ and a
vector is normalized if $\mathbf{V}^{\dagger}\cdot\mathbf{V}=1.$ The outer
product of an $n\times1$ vector $\mathbf{V}$ is the $n\times n$ matrix
$\mathbf{V}\cdot\mathbf{V}^{\dagger}$ with $v_{i}\cdot v_{j}^{\ast}$ in row
$i$ column $j$. If $M$ is an $n\times m$ matrix with value $v_{ij}$ in row $i$
and column $j$, then the Hermitian transpose, $M^{\dagger}$, is a $m\times n$
matrix with value $v_{ji}^{\ast} $ in row $i$ and column $j$. A projector $M$
is a matrix that is Hermitian and idempotent, $M=M^{\dagger}=M\cdot M$. If
$A"$ is a subspace that is spanned by eigenvectors $\{V_{j}$, $j\in A"\}$,
then the projector for this subspace equals the sum of outer products
$M_{A}=\sum_{j\in A"}$\ $\mathbf{V}_{j}\cdot\mathbf{V}_{j}^{\dagger}$. A
unitary matrix is a square matrix $T$ that satisfies $T^{\dagger
}T=I=TT^{\dagger}$, which is property that is needed to preserve the lengths
of vectors. The quantum probability of an event $A"$ can be expressed as
follows:%
\begin{gather}
q(A")=\left\vert M_{A}\cdot\mathbf{\psi}\right\vert ^{2}=\left\vert \sum_{j\in
A"}\mathbf{V}_{j}\cdot\mathbf{V}_{j}^{\dagger}\cdot\mathbf{\psi}\right\vert
^{2}=\left\vert \sum_{j\in A"}\mathbf{V}_{j}\cdot\psi_{j}\right\vert
^{2}\label{Qprob}\\
\text{and because of orthogonality of eigenvectors}\nonumber\\
=\sum\nolimits_{j\in A"}\left\vert \mathbf{V}_{j}\cdot\psi_{j}\right\vert
^{2}=\sum\nolimits_{j\in A"}\left\vert \mathbf{V}_{j}\right\vert ^{2}%
\cdot\left\vert \psi_{j}\right\vert ^{2}\nonumber\\
\text{and because eigenvectors are unit length}\nonumber\\
=\sum\nolimits_{j\in A"}1\cdot\left\vert \psi_{j}\right\vert ^{2}%
=\sum\nolimits_{j\in A"}\left\vert \psi_{j}\right\vert ^{2}.\nonumber
\end{gather}

\subsection{Compatible events}

First we prove that if (a) the $q$ events in $Q=\{X",Y",Z",...\}$ are mutually
exclusive and exhaustive, (b) the $r$ events in $R=\{U",V",W",...\}$ are
mutually exclusive and exhaustive, (c) the events in $Q$ are not orthogonal to
the events in $R$, but (d) the events in $Q$ are all compatible with the
events in $R,$ then we require at least a $q\cdot r$-dimensional vector space.
\ Assumption (a) implies $M_{i}\cdot M_{i^{\prime}}=0$ for $i\neq i^{\prime}%
$and $\sum_{i}M_{i}=I$ for events selected from $Q$; assumption (b) implies
$M_{j}\cdot M_{j^{\prime}}=0$ for $j\neq j^{\prime}$and $\sum M_{j}=I$ \ for
events selected from $R$; and assumption (c) implies $M_{i}\cdot M_{j}\neq0$
for $i\in Q$ and $j\in R$; and assumption (d) implies $M_{i}\cdot M_{j}%
=M_{j}\cdot M_{i}$ for any pair of events. Each matrix product $M_{ij}%
=M_{i}\cdot M_{j}$ for $i\in Q$ and $j\in R$ is a projector because
$(M_{i}\cdot M_{j})^{\dagger}=M_{j}\cdot M_{i}=(M_{i}\cdot M_{j})$ and
$(M_{i}M_{j})\cdot(M_{i}M_{j})=(M_{j}M_{i}M_{i}M_{j})$ $=(M_{j}M_{i}%
M_{j})=(M_{j}M_{j}M_{i})=$ $M_{j}M_{i}=(M_{i}\cdot M_{j})$. Also each matrix
product $M_{ij}=M_{i}\cdot M_{j}$ for $i\in Q$ and $j\in R$ projects on to at
least one dimension because it is a nonzero matrix ($M_{i}\cdot M_{j}\neq0)$.
Finally each pair of matrix products is orthogonal because $(M_{i}M_{j}%
)\cdot(M_{i^{\prime}}M_{j^{\prime}})=(M_{i}M_{j}M_{j^{\prime}}M_{i^{\prime}%
})=(M_{j}M_{i}M_{i^{\prime}}M_{j^{\prime}})=0$ if $i\neq i^{\prime}$ or $j\neq
j^{\prime}$ for $i,i^{\prime}\in Q$ and $j,j^{\prime}\in R$. Thus each matrix
product $M_{ij}=M_{i}\cdot M_{j}$ is a projector that projects onto an
orthogonal subspace associated with the intersection of events $i"\cap j"$.
Finally note that $I=I\cdot I=$ $\left(  \sum_{i\in Q}M_{i}\right)
\cdot\left(  \sum_{j\in R}M_{i}\right)  =\sum_{i}\sum_{j}M_{ij} $, and so the
product matrices $M_{ij}=M_{i}\cdot M_{j}$ for $i\in Q$ and $j\in R$ form a
spectral resolution of the identity. The identity projects onto the entire
vector space, and so the dimension of the vector space equals $Rank(I)$
$=Rank(\sum_{i}\sum_{j}M_{ij})=\sum_{i}\sum_{j}rank(M_{ij})$, for orthogonal
projectors $M_{ij}$. If each product matrix has rank one (the minimum for a
nonzero projector), then each product matrix has one eigenvector,
$\mathbf{V}_{ij}$, and the product matrix can be computed from the outer
product of its eigenvector $M_{ij}=\mathbf{V}_{ij}\cdot\mathbf{V}%
_{ij}^{\dagger}$. In this case, any vector in the vector space can be
described as linear combination of the eigenvectors $\mathbf{V}_{ij}$
representing the joint event $i"\cap j"$for $i\in Q$ and $j\in R$:%
\begin{align*}
\mathbf{\psi}  & =I\cdot\mathbf{\psi}\\
& =\sum_{i}\sum_{j}M_{ij}\cdot\mathbf{\psi}\\
& =\sum_{i}\sum_{j}\mathbf{V}_{ij}\cdot\mathbf{V}_{ij}^{\dagger}%
\cdot\mathbf{\psi}\\
& =\sum_{i}\sum_{j}(\mathbf{V}_{ij}^{\dagger}\cdot\mathbf{\psi})\cdot
\mathbf{V}_{ij}.
\end{align*}
This vector space, expressed in terms of the `joint' eigenvectors
$\mathbf{V}_{ij}$ is also described as a tensor product space.

Second, we prove that if events $A",B"$ are compatible, then the sequential
conjunction obeys the commutative and distributive rules. $(A"\sqcap B")$ is
true if and only if the final projection is contained in the subspace
corresponding to the projector $M_{B}\cdot M_{A}$; $(B"\sqcap A")$ is true if
and only if the final projection is contained in the subspace corresponding to
the projector $M_{A}\cdot M_{B}$; but these two projectors are identical
because they commute, $M_{B}\cdot M_{A}=M_{A}\cdot M_{B}.$ Furthermore,
$M_{B}\cdot M_{A}=M_{A}\cdot M_{B}=M_{A\cap B}$ is the projector for the
subspace spanned by eigenvectors that are common between the two events, which
equals the intersection of the two subspaces, $A"\cap B"$. Thus $(A"\sqcap
B")=$ $(A"\cap B")$ for compatible events. $A"\sqcap(B"\sqcup B^{\perp})$ is
true if and only if $A"$ is true because $(B"\sqcup B^{\perp})$ is always
true; $(A"\sqcap B")$ is true if and only if the final projection is contained
in the intersection $(A"\cap B")$; $(A"\sqcap B^{\perp})$ is true if and only
if the final projection is contained in the intersection $(A"\cap B^{\perp})$;
finally $(A"\sqcap B")$ $\sqcup$ $(A"\sqcap B^{\perp})$ is true if and only if
the final projection is contained in $(A"\cap B")$ or $(A"\cap B^{\perp})$,
and the latter is true if and only if the final projection is contained in
$(A"\cap B")$ $\cup$ $(A"\cap B^{\perp})=$ $A"$.

Third, we prove that if $M_{A}$ and $M_{B}$ commute, then the quantum
probabilities obey the classic probability (Kolmogorov) rules. Immediately
above we proved that if $M_{A}$ and $M_{B}$ commute, then $A"\sqcap B"=$
$A"\cap B"$ and therefore $q(A"\sqcap B")=q(A"\cap B")=q(B"\cap A").$ From
this it also follows that%
\[
q(A"\sqcup B")=1-q(A^{\perp}\sqcap B^{\perp})=1-q(A^{\perp}\cap B^{\perp}).
\]

and%

\begin{align*}
q(A"|B")  & =|M_{A}\mathbf{\psi}_{B}|^{2}\\
& =|M_{A}M_{B}\mathbf{\psi}|^{2}/|M_{B}\mathbf{\psi}|^{2}\\
& =q(A"\cap B")/q(A").
\end{align*}
If the events are commutative, then the law of total probability also holds
\begin{align*}
q(A")  & =q(A"\sqcap(B\sqcup B^{\perp}))\\
& =q(A")q(B"|A")+q(A")q(B^{\perp}|A")\\
& =q(A"\sqcap B")+q(A"\sqcap B^{\perp})\\
& =q(A"\cap B")+q(A"\cap B^{\perp})\\
& =q(B"\cap A")+q(B^{\perp}\cap A").
\end{align*}

\subsection{Derivation for interference terms}

Next we derive Equation \ref{QConjEff}.%

\begin{align}
q(L")  & =\left\vert M_{L}\mathbf{\psi}\right\vert ^{2}=\left\vert M_{L}\cdot
I\cdot\mathbf{\psi}\right\vert ^{2}=\left\vert M_{L}\cdot(M_{H}+M_{H^{\perp}%
})\cdot\mathbf{\psi}\right\vert ^{2}\\
& =\mathbf{\psi}^{\dagger}(M_{H}+M_{H^{\perp}})\cdot M_{L}\cdot M_{L}%
\cdot(M_{H}+M_{H^{\perp}})\mathbf{\psi}\nonumber\\
& =\mathbf{\psi}^{\dagger}(M_{H}+M_{H^{\perp}})\cdot M_{L}\cdot(M_{H}%
+M_{H^{\perp}})\mathbf{\psi}\nonumber\\
& =\mathbf{\psi}^{\dagger}(M_{H}M_{L}M_{H}+M_{H^{\perp}}M_{L}M_{H}+M_{H}%
M_{L}M_{H^{\perp}}+M_{H^{\perp}}M_{L}M_{H^{\perp}})\mathbf{\psi}\nonumber\\
& =q(H"\sqcap L")+[\mathbf{\psi}^{\dagger}M_{H^{\perp}}M_{L}M_{H}\mathbf{\psi
}+\mathbf{\psi}^{\dagger}M_{H}M_{L}M_{H^{\perp}}\mathbf{\psi}]+q(H^{\perp
}\sqcap L")\nonumber\\
& =q(H"\sqcap L")+Int_{L}+q(H^{\perp}\sqcap L").\nonumber
\end{align}
Further analysis of the interference proves that
\begin{align*}
Int_{L}  & =\mathbf{\psi}^{\dagger}M_{H^{\perp}}M_{L}M_{H}\mathbf{\psi
}+\mathbf{\psi}^{\dagger}M_{H}M_{L}M_{H^{\perp}}\mathbf{\psi}\\
& =\left(  \mathbf{\psi}^{\dagger}M_{H^{\perp}}M_{L}M_{H}\mathbf{\psi}\right)
+\left(  \mathbf{\psi}^{\dagger}M_{H^{\perp}}M_{L}M_{H}\mathbf{\psi}\right)
^{\dagger}\\
& =\left(  \mathbf{\psi}^{\dagger}M_{H^{\perp}}M_{L}M_{H}\mathbf{\psi}\right)
+\left(  \mathbf{\psi}^{\dagger}M_{H^{\perp}}M_{L}M_{H}\mathbf{\psi}\right)
^{\ast}\\
& =2\cdot\operatorname{Re}[\mathbf{\psi}^{\dagger}M_{H^{\perp}}M_{L}%
M_{H}\mathbf{\psi}]\\
& =2\cdot\operatorname{Re}[(M_{L}M_{H^{\perp}}\mathbf{\psi})^{\dagger}%
(M_{L}M_{H}\mathbf{\psi})].
\end{align*}
It is useful to consider the simple case in which the vector space is 2
dimensional and all events are rays. Then $M_{j}=\mathbf{V}_{j}\mathbf{V}%
_{j}^{\dagger}$ and the interference term reduces to the special case used in
\cite{Franco_JMP_2009}:%

\begin{align*}
& 2\cdot\operatorname{Re}[(M_{L}M_{H^{\perp}}\mathbf{\psi})^{\dagger}%
(M_{L}M_{H}\mathbf{\psi})]\\
& =2\cdot\operatorname{Re}[(\mathbf{V}_{L}\mathbf{V}_{L}^{\dagger}%
\mathbf{V}_{H^{\perp}}\mathbf{V}_{H^{\perp}}^{\dagger}\mathbf{\psi})^{\dagger
}(\mathbf{V}_{L}\mathbf{V}_{L}^{\dagger}\mathbf{V}_{H}\mathbf{V}_{H}^{\dagger
}\mathbf{\psi})]\\
& =2\cdot\operatorname{Re}[(\mathbf{V}_{L}(\mathbf{V}_{L}^{\dagger}%
\mathbf{V}_{H^{\perp}})(\mathbf{V}_{H^{\perp}}^{\dagger}\mathbf{\psi
)})^{\dagger}(\mathbf{V}_{L}(\mathbf{V}_{L}^{\dagger}\mathbf{V}_{H}%
)(\mathbf{V}_{H}^{\dagger}\mathbf{\psi}))]\\
& =2\cdot\operatorname{Re}[(\mathbf{V}_{L}^{\dagger}\mathbf{V}_{H^{\perp}%
})^{\ast}\cdot(\mathbf{V}_{H^{\perp}}^{\dagger}\mathbf{\psi})^{\ast}%
\cdot(\mathbf{V}_{L}^{\dagger}\mathbf{V}_{L})\cdot(\mathbf{V}_{L}^{\dagger
}\mathbf{V}_{H})\cdot(\mathbf{V}_{H}^{\dagger}\mathbf{\psi})]\\
& =2\cdot\operatorname{Re}[(\mathbf{V}_{L}^{\dagger}\mathbf{V}_{H^{\perp}%
})^{\ast}\cdot(\mathbf{V}_{H^{\perp}}^{\dagger}\mathbf{\psi})^{\ast}%
\cdot(1)\cdot(\mathbf{V}_{L}^{\dagger}\mathbf{V}_{H})\cdot(\mathbf{V}%
_{H}^{\dagger}\mathbf{\psi})]\\
& =2\cdot\operatorname{Re}[(\mathbf{V}_{L}^{\dagger}\mathbf{V}_{H^{\perp}%
})^{\ast}\cdot(\mathbf{V}_{H^{\perp}}^{\dagger}\mathbf{\psi})^{\ast}%
\cdot(\mathbf{V}_{L}^{\dagger}\mathbf{V}_{H})\cdot(\mathbf{V}_{H}^{\dagger
}\mathbf{\psi})]
\end{align*}

If the events are compatible, then the interference is zero:
\begin{align*}
Int_{L}  & =2\cdot\operatorname{Re}[(M_{L}M_{H^{\perp}}\mathbf{\psi}%
)^{\dagger}(M_{L}M_{H}\mathbf{\psi})]\\
& 2\cdot\operatorname{Re}[\mathbf{\psi}^{\dagger}M_{H^{\perp}}M_{L}%
M_{H}\mathbf{\psi}]\\
& =2\cdot\operatorname{Re}[\mathbf{\psi}^{\dagger}\cdot(M_{H^{\perp}}%
M_{H})\cdot M_{L}\mathbf{\psi}]\\
& =2\cdot\operatorname{Re}[\mathbf{\psi}^{\dagger}\cdot0\cdot M_{L}%
\mathbf{\psi}]=0.
\end{align*}

If $q(A")>q(B")$, then a conjunction error can only occur for the lower
probability event, no matter what order is processed:%

\begin{align*}
q(A")  & \geq q(A")\cdot q(B"|A")=q(A"\sqcap B"),\\
q(A")  & >q(B")\geq q(B")\cdot q(A"|B")=q(B"\sqcap A"),
\end{align*}

If $q(A)=q(B)$, then there can be no conjunction error:
\begin{align*}
q(B")  & =q(A")\geq q(A")\cdot q(B"|A")=q(A"\sqcap B"),\\
q(A")  & =q(B")\geq q(B")\cdot q(B"|A")=q(B"\sqcap A").
\end{align*}

Now consider the relation between $Int_{L}$ and $Int_{L^{\perp}}$. \ First we
note that if we query $H$ first, then $q(H")=q(H"\sqcap L")+q(H"\sqcap
L^{\perp})$, producing no interference for $H$, so%
\begin{align*}
1  & =q(L")+q(L^{\perp})\\
& =[q(H"\sqcap L")+q(H^{\perp}\sqcap L")+Int_{L}]+[q(H"\sqcap L^{\perp
})+q(H^{\perp}\sqcap L^{\perp})+Int_{L^{\perp}}]\\
& =q(H"\sqcap L")+q(H"\sqcap L^{\perp})+q(H^{\perp}\sqcap L")+q(H^{\perp
}\sqcap L^{\perp})+(Int_{L}+Int_{L^{\perp}})\\
& =[q(H"\sqcap L")+q(H"\sqcap L^{\perp})]+[q(H^{\perp}\sqcap L")+q(H^{\perp
}\sqcap L^{\perp})]\\
& =q(H")+q(H^{\perp})=1.
\end{align*}
These equalities imply that $Int_{L}+Int_{L^{\perp}}=0.$ The same argument
holds for $Int_{H}+Int_{H^{\perp}}=0.$

For the next property, it is useful to express the interference as follows:%

\begin{align*}
Int_{L}  & =2\cdot\operatorname{Re}[\mathbf{\psi}^{\dagger}M_{H^{\perp}}%
M_{L}M_{H}\mathbf{\psi}]\\
& =2\cdot\operatorname{Re}[\mathbf{\psi}^{\dagger}(I-M_{H})M_{L}%
M_{H}\mathbf{\psi}]\\
& =2\cdot\operatorname{Re}[\mathbf{\psi}^{\dagger}(M_{L}M_{H}-M_{H}M_{L}%
M_{H})\mathbf{\psi}]\\
& =2\cdot\operatorname{Re}[\mathbf{\psi}^{\dagger}M_{L}M_{H}\mathbf{\psi
}]-2\cdot\mathbf{\psi}^{\dagger}M_{H}M_{L}M_{H}\mathbf{\psi}\\
& =2\cdot\operatorname{Re}[\mathbf{\psi}^{\dagger}M_{L}M_{H}\mathbf{\psi
}]-2\cdot q(H"\sqcap L")\\
& =2\cdot\{\operatorname{Re}[(M_{L}\mathbf{\psi})^{\dagger}\cdot
(M_{H}\mathbf{\psi})]-q(H"\sqcap L")\}.
\end{align*}

Satisfying both the conjunction and disjunction errors implies the following
inequalities. The conjunction error requires $Int_{L}<-q(H^{\perp}\sqcap L") $
and the disjunction error requires $Int_{H^{\perp}}<-q(L"\sqcap H^{\perp})$
and the latter implies $Int_{H}>q(L"\sqcap H^{\perp}).$We know that
$Int_{H}=2\cdot\{\operatorname{Re}[\left(  M_{H}\mathbf{\psi}\right)
^{\dagger}\left(  M_{L}\mathbf{\psi}\right)  ]-q(L"\sqcap H")\}.$ This implies
that we must satisfy the following constraints:%

\begin{align*}
Int_{H}  & =2\cdot\operatorname{Re}[\left(  M_{H}\mathbf{\psi}\right)
^{\dagger}\left(  M_{L}\mathbf{\psi}\right)  ]-2\cdot q(L"\sqcap H")\\
& >q(L"\sqcap H^{\perp})>-q(H^{\perp}\sqcap L")\\
& >2\cdot\operatorname{Re}[\left(  M_{L}\mathbf{\psi}\right)  ^{\dagger
}\left(  M_{H}\mathbf{\psi}\right)  ]-2\cdot q(H"\sqcap L")=Int_{L}%
\end{align*}
Given the fact that $\operatorname{Re}[\left(  M_{H}\mathbf{\psi}\right)
^{\dagger}\left(  M_{L}\mathbf{\psi}\right)  ]=\operatorname{Re}[\left(
M_{L}\mathbf{\psi}\right)  ^{\dagger}\left(  M_{H}\mathbf{\psi}\right)  ]$, we
see that this set of constraints requires $q(L"\sqcap H")=\left\vert
M_{H}M_{L}\mathbf{\psi}\right\vert ^{2}$ $<\left\vert M_{L}M_{H}\mathbf{\psi
}\right\vert ^{2}=q(H"\sqcap L")$ which is consistent with the theory when the
events are incompatible.

\end{document}